\documentclass[reprint,amsmath,amssymb,aps,prb]{revtex4-2}
\usepackage{graphicx} 
\usepackage{subcaption} 
\usepackage[labelsep=period,font=small,labelsep=period]{caption} 
\usepackage{ragged2e} 
\usepackage{float} 

\usepackage{latexsym, amsmath,amssymb,bm,euscript} 
\usepackage{xcolor} 
\usepackage[colorlinks=true,linkcolor=blue,citecolor=blue,urlcolor=blue]{hyperref}
\usepackage{ulem}

\usepackage[T1]{fontenc}
\usepackage{lmodern}
\begin{document}

\title{Emergent Fermi-liquid-like phase by melting a holon Wigner crystal in a doped Mott insulator on the kagome lattice}

\author{Xu-Yan Jia$^1$}
\author{Wen Huang$^{2,3}$}
\author{D. N. Sheng$^4$}
\email{donna.sheng1@csun.edu}
\author{Shou-Shu Gong$^{2,3}$}
\email{shoushu.gong@gbu.edu.cn}

\affiliation{$^1$School of Physics, Beihang University, Beijing 100191, China}
\affiliation{$^2$School of Physical Sciences, Great Bay University, Dongguan 523000, China}
\affiliation{$^3$Great Bay Institute for Advanced Study, Dongguan 523000, China}
\affiliation{$^4$Department of Physics and Astronomy, California State University Northridge, Northridge, California 91330, USA}

\date{\today}

\begin{abstract}
The doped quantum spin liquid on the kagome lattice provides a fascinating platform to explore exotic quantum states, such as the reported holon Wigner crystal at low doping.
By extending the doping range to $\delta = 0.027$ – $0.36$, we study the kagome-lattice $t$–$J$ model using the state-of-the-art density matrix renormalization group calculation.
On the $L_y=3$ cylinder ($L_y$ is the number of unit cells along the circumference direction), we establish a quantum phase diagram with increasing doping level.
In addition to the charge density wave (CDW) states at lower doping, we find an emergent Fermi-liquid-like phase by melting the holon Wigner crystal at $\delta \approx 0.15$, which is characterized by suppression of charge density oscillation and power-law decay of various correlation functions. 
On the wider $L_y = 4$ cylinder, the bond-dimension extrapolated correlation functions also support such a Fermi-liquid-like state, suggesting its stability with increasing system size.
In a narrow doping range near $\delta = 1/3$ on the $L_y = 3$ cylinder, we find a state with an exponential decay of single-particle correlation but the other correlation functions preserving the features in the Fermi-liquid-like phase, which may be a precursor of a superconducting state. 
Nevertheless, this peculiar state near $\delta = 1/3$ disappears on the $L_y = 4$ cylinder, implying a possible lattice size dependence.
Our results reveal a quantum melting from a holon Wigner crystal to a Fermi-liquid-like state with increasing hole density, and suggest a doping regime to explore superconductivity for future study.
\end{abstract}

\maketitle

\section{Introduction}
With the interplay among band topology, geometry frustration, and electronic correlation, kagome-lattice systems realize a fascinating platform to explore exotic quantum states and phenomena~\cite{kagome-RV-nature_reviews_physics-2023,kagome-RV-nature-2022,kagome-AV3Sb5-National_Science_Review-2022,kagome-AV3Sb5-nature_Review_material-2024,kagome-Co3Sn2S2-Reviews_in_physics-2022,kagome-Co3Sn2S2-sci-2019-2,kagome-Co3Sn2S2-sci-2019,kagome-CsV3Sb5-nature-2021,kagome-CsV3Sb5-nature-2021-2,kagome-Mn3Sn-nature-2015,kagome-TbMn6Sn6-nature-2020}. 
Among the various topics, an important subject is superconductivity (SC).
Besides the intensively studied kagome superconductors $\mathrm{AV_3Sb_5(A=K,Rb,Cs)}$~\cite{kagome-RV-nature-2022,kagome-AV3Sb5-National_Science_Review-2022,kagome-AV3Sb5-nature_Review_material-2024,kagome-CsV3Sb5-nature-2021,kagome-CsV3Sb5-nature-2021-2}, in which the Fermi level lies close to a Van Hove singularity, another route to SC is to dope a quantum spin liquid (QSL) or an antiferromagnetic Mott insulator on the kagome lattice~\cite{savary_2017,zhou_2017,broholm_2020}.
In the study of cuprate superconductors, Anderson proposed that doping a QSL parent state can naturally give rise to an unconventional SC~\cite{anderson_1987}. 
Furthermore, it was conjectured that doping a chiral spin liquid may even lead to a topological SC through the condensation of paired fractional quasiparticles~\cite{CSL-Laughlin-1988,CSL-xiaogang-1989,CSL-Fisher-1989}.
Recently, there has been unbiased numerical evidence to support superconducting states in doped Mott insulators on the square~\cite{Square-tt'JJ'-JiangHC-2021,Square-tt'JJ'-gss-2021,Square-tt'J-JiangShengtao-2021,Square-tt'JJ'-gss-2024,chen_2025,jiang_2023,jiang_2024} and triangular~\cite{jiang_2020,huang_2022,triangular-tt'JJ'-HuangYX-2023} lattices.

For kagome systems, many antiferromagnets have been proposed as candidates for QSL in experiment~\cite{kagome-Herbertsmithite-RMP-2016,savary_2017,zhou_2017,broholm_2020}.
In model study, the spin-$1/2$ kagome model with the nearest-neighbor (NN) Heisenberg interaction has been identified to host a QSL ground state, which may be stabilized by geometric frustration and a low coordination number.
However, the nature of this state is not fully understood, and there is a debate between a gapless $U(1)$ QSL and a gapped $Z_2$ QSL~\cite{kagome-J1-VMC-2007,kagome-J1-VMC-2011,kagome-J1-VMC-2013,kagome-J1-VMC-2014,kagome-J1-DMRG-Sci-2011,jiang_2008,kagome-J1-DMRG-PRL-2012,liao_2017,kagome-J1-iDMRG-2017,lauchli_2019,kagome-J1J2J3-mean_field-2012,zhu_2018,sun_2024,pinaki_2025,kagome_spin_model_zw}. 
Beyond the NN Heisenberg interaction, it is remarkable that some additional interactions can drive a Kalmeyer-Laughlin chiral spin liquid with a topological spin Chern number~\cite{kalmeyer_1987}, such as further-neighbor antiferromagnetic interactions~\cite{kagome-J1J2szJ3sz-heyinchen-2014,kagome-J1J2J3-gss-2014,kagome-J1J2J3-ssg-2015} or three-spin scalar chiral coupling~\cite{kagome-J1Jchihz-Ncom-2014,kagome-J1J2J3/J1Jchi-VMC-2015,reza2019}.

The discovered QSL states in kagome systems quickly stimulated interest in the doped case.
Early variational Monte Carlo (VMC) studies on the kagome $t$-$J$ model found that even a small amount of hole doping may drive a zero-flux state with a valence bond crystal (VBC) order, which persists up to the doping level of $\delta \approx 0.18$~\cite{kagome-tJ-VMC-2011,kagome-tJ-VMC-2013}.
Later, by considering the SU(2)-gauge rotation in the mean-field Hamiltonian, the VMC study obtained a chiral noncentrosymmetric nematic superconducting state at $\delta \lesssim 0.02$, and the VBC state is restored at higher doping~\cite{kagome-tJ-VMC-2021}.
On the other hand, the density matrix renormalization group (DMRG) calculations up to $\delta \approx 0.11$ in finite-width cylinder systems obtained an insulating charge density wave (CDW) phase~\cite{kagome-tJ-Jiang-2017}, manifesting itself as a stripe crystal or a Wigner crystal depending on doping concentration and cylinder geometry.
Because of the short-range spin correlation which is similar to the QSL parent state, this CDW phase was understood as a crystalline state formed by spinless holons instead of holes~\cite{kagome-tJ-Jiang-2017,kagome-tJlike-PCheng-2021,kivelson1987}.
With an additional chiral coupling, the doped chiral spin liquid also gives rise to these CDW states~\cite{kagome-tJlike-PCheng-2021}. 
Interestingly, in the lightly doped extended $t$–$J$ model, the increasing of next-nearest-neighbor couplings can melt the CDW states and lead to a Fermi-liquid–like phase~\cite{extend-kagome-tj}.

At higher doping level, many exotic charge uniform states have been theoretically proposed in doped QSLs, such as holon metal, fractionalized Fermi liquid, and SC~\cite{law2017}.
Recently, a study employing the projected entangled simplex states (PESS) provides new results for the kagome $t$-$J$ model in a wider doping range~\cite{kagome-tJ-GuZC-2024}.
The PESS simulations find different types of Wigner crystal depending on the doping level for $\delta \lesssim 0.27$.
For the higher doping concentration ($0.27 < \delta < 1/3$), the PESS study obtains uniform states, including a non-Fermi liquid state characterized by an exponential decay of single-particle and spin correlation functions ($0.27 < \delta < 0.32$), and a pair density wave state ($0.32 < \delta < 1/3$).
These novel findings have sparked growing interest in the kagome-lattice $t$-$J$ model under higher doping.

\begin{figure}[t]
	\centering
	\begin{subfigure}[b]{0.48\textwidth}
		\raggedleft
		\includegraphics[width=0.85\textwidth]{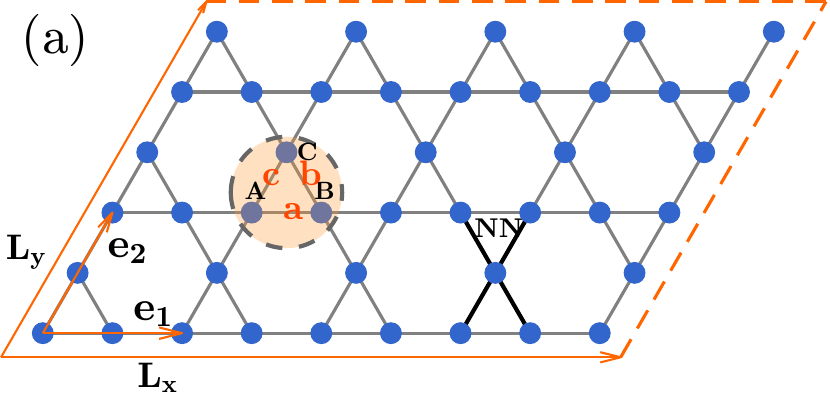} 
	\end{subfigure}
    \begin{subfigure}[b]{0.48\textwidth}
	   \includegraphics[width=\textwidth]{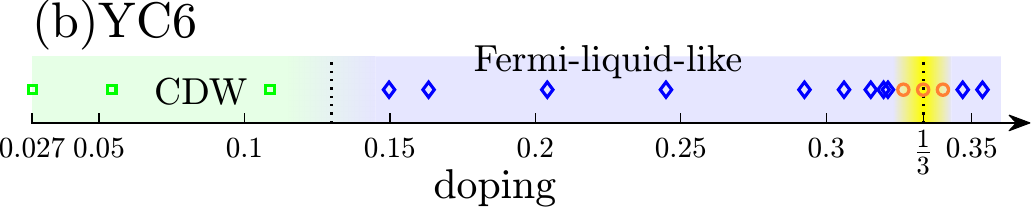}
    \end{subfigure}
	\caption{\justifying Schematic figure of kagome-lattice $t$-$J$ model and quantum phase diagram of the YC6 system with increasing doping ratio $\delta = 0.027 - 0.36$. (a) The $t$-$J$ model on the kagome Y-cylinder (YC), where the periodic and open boundary conditions are imposed, respectively, along the directions specified by the lattice vectors ${\bf e}_2$ and ${\bf e}_1$. The electrons and doped holes reside at lattice vertices (solid circles). Each unit cell (denoted by the small triangle in the shaded region) has three sites ($A$, $B$, and $C$) and three bonds ($a$, $b$, and $c$). $L_x$ and $L_y$ are the numbers of unit cells along the ${\bf e}_1$ and ${\bf e}_2$ directions, respectively. Note that the column and row indices (x, y) will be used later to specify the coordinates of individual lattice sites. In the example illustrated in the figure, sites $A$ and $B$ correspond to coordinates $(3, 3)$ and $(4, 3)$, respectively. (b) Quantum phase diagram of the kagome-lattice $t$-$J$ model ($t/J = 3$) in the YC6 ($L_y = 3$) system. Besides the charge density wave (CDW) states observed previously~\cite{kagome-tJ-Jiang-2017,kagome-tJlike-PCheng-2021}, we identify a new Fermi-liquid-like phase. In a narrow regime near $1/3$ doping, the state shows an exponential decay of single-particle correlation but the other correlation functions remain the features in the Fermi-liquid-like phase.} 
    \label{lattice_diagram}
\end{figure}

In this work, we employ the DMRG calculation to investigate the kagome-lattice $t$-$J$ model in a wide doping range of $\delta = 0.027 - 0.36$. 
We focus on the cylindrical geometry with the circumference of $3$ unit cells ($L_y = 3$) and map a phase diagram with increasing doping level, as shown in Fig.~\ref{lattice_diagram}.
At lower doping, we confirm the CDW states that have been reported in previous DMRG calculations~\cite{kagome-tJ-Jiang-2017,kagome-tJlike-PCheng-2021}. 
At $\delta \approx 0.15$, the CDW order is melted and the measured correlation functions exhibit a substantial enhancement. 
For $0.15 \lesssim \delta \lesssim 0.36$ (except a narrow regime near $\delta = 1/3$), we identify a Fermi-liquid-like phase characterized by the power-law decay of the measured correlation functions and the absence of hole pairing.
In a very narrow doping range near $\delta = 1/3$, the single-particle correlation becomes an exponential decay, but the other correlation functions preserve the features in the Fermi-liquid-like phase.
In particular, the SC pairing correlation function is much stronger than the square of single-particle correlation, which characterizes the hole pairing and thus suggests this state as a possible precursor of a superconducting state like the pseudogap state. 
On the wider $L_y = 4$ system, the Fermi-liquid-like state is supported by the correlation functions, but the emergent state near $1/3$ doping disappears, suggesting that this phase may have a lattice size dependence.
In the summary section, we discuss future studies on the phase transition of melting the holon Wigner crystal, and the potential routes to search for the SC near $\delta = 1/3$ by considering other interactions.

\section{MODEL AND METHOD}

The Hamiltonian of the kagome-lattice $t$-$J$ model is given as
\begin{align*}
 H \! = -t\!\!\sum\limits_{\langle ij \rangle,\sigma}\!(\hat{c}_{i,\sigma}^\dagger \hat{c}_{j,\sigma}+\mathrm{H.c.})\!+\!J\sum\limits_{\langle ij \rangle}\!(\hat{\bf S}_i\cdot \hat{\bf S}_j\!-\!\frac{1}{4}\hat{n}_{i} \hat{n}_{j}),
\end{align*}
where $\hat{c}^\dagger_{i,\sigma}$ and $\hat{c}_{i,\sigma}$ are the creation and annihilation operators for an electron with spin $\sigma$ ($\sigma = \pm 1/2$) at the site $i$, $\hat{\bf S}_i$ is the spin-$1/2$ operator, and $\hat{n}_i \equiv \sum_\sigma \hat{c}^\dagger_{i,\sigma} \hat{c}_{i,\sigma}$ is the electron number operator. 
The $t$-$J$ model enforces the no-double-occupancy constraint at each site.
In this work, we consider the system with the NN couplings $t$ and $J$ in the doping range $\delta = 0.027 - 0.36$, and we set $t/J = 3$ to mimic the large Hubbard-$U$ case.

To determine the ground state of this model, we perform DMRG~\cite{DMRG-White-1992} calculations on the cylindrical lattice, which is illustrated in Fig.~\ref{lattice_diagram}(a), where $\mathbf{e}_1$ and $\mathbf{e}_2$ denote the two unit vectors.  
We employ the kagome Y-cylinder (YC) with periodic (open) boundary conditions along the $\mathbf{e}_2$ ($\mathbf{e}_1$) direction. 
We denote the number of unit cells along the $\mathbf{e}_2$ and $\mathbf{e}_1$ directions as $L_y$ and $L_x$, respectively. 
Note that we add an extra column of lattice sites at the right boundary, where each unit cell only contains $A$ and $C$ sites, as shown in Fig.~\ref{lattice_diagram}(a). 
This YC cylinder with the modified boundary conditions can reduce the open boundary effects.
Therefore, the lattice has $2L_x + 1$ columns and $2L_y$ rows, with the column index ranging as $x \in [1, 2L_x+1]$ and the row index as $y \in [1, 2L_y]$.
We denote the cylinders as YC$2L_y$, with the total site number $N = 3 L_y L_x +2L_y$. 
The doping ratio $\delta$ is defined as $\delta = N_h / N$, where $N_h$ is the number of doped holes.
We mainly study the YC6 cylinder ($L_y = 3$) with $L_x = 32$, and we also examine the YC8 cylinder for some typical parameter points.
By combining the charge $U(1)$ and spin $SU(2)$ symmetries~\cite{McCulloch_2002}, we keep the bond dimensions up to $D = 28000$ $SU(2)$ multiplets, which are equivalent to about $84000$ $U(1)$ states. 
For all parameter points in this work, the DMRG truncation error of the largest bond dimension is controlled below $2 \times 10^{-7}$, giving us accurate results.

\section{Charge density wave and Fermi-liquid-like phase}
In Fig.~\ref{lattice_diagram}(b), we present the phase diagram of the YC6 system. 
In the doping range $\delta = 0.027 - 0.11$, previous DMRG studies have found insulating CDW states~\cite{kagome-tJ-Jiang-2017,kagome-tJlike-PCheng-2021}. 
In this work, we further investigate the higher doping level and identify a Fermi-liquid-like phase in the doping range $\delta \approx 0.15$–$0.36$, with the exception of $\delta \approx 1/3$.

The CDW and the Fermi-liquid-like phases can be distinguished by their charge density distributions and correlation functions. 
In the CDW phase, the charge density distribution shows a striped crystal or a Wigner crystal depending on the lattice geometry and doping ratio~\cite{kagome-tJ-Jiang-2017,kagome-tJlike-PCheng-2021}.
The CDW states have a stable charge density oscillation with a relatively large amplitude.
The superconducting, single-particle and spin-spin correlation functions demonstrate an exponential decay in this phase, with very small correlation lengths~\cite{kagome-tJ-Jiang-2017,kagome-tJlike-PCheng-2021}.
Upon increasing the doping level, the emergent Fermi-liquid-like phase displays strikingly different characteristics. 
The charge density distribution becomes uniform along the $\mathbf{e}_2$ direction, and along the $\mathbf{e}_1$ direction it shows a very small oscillation amplitude characterizing the melted static CDW order. 
Moreover, the correlation functions display the power-law rather than exponential decay. 
In particular, the magnitude of the SC pairing correlation is comparable to the square of single-particle correlation, indicating the absence of hole pairing.
The finite central charge obtained by fitting the entanglement entropy also supports its gapless nature.
These features consistently characterize this new phase as a Fermi-liquid-like state.

\subsection{Charge density profiles on YC6}
\begin{figure}[h] 
	\centering
	\begin{subfigure}[b]{0.48\textwidth}
		\includegraphics[width=\textwidth]{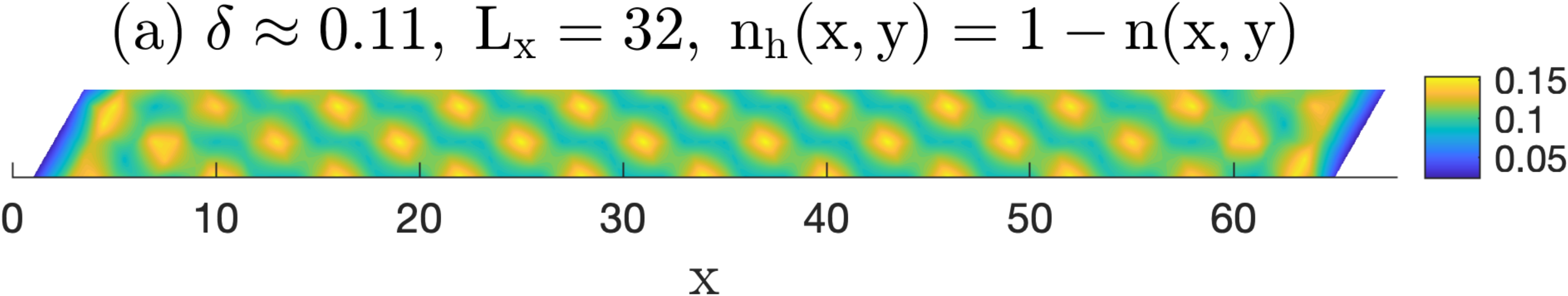} 
	\end{subfigure}
	\begin{subfigure}[b]{0.48\textwidth}
		\includegraphics[width=\textwidth]{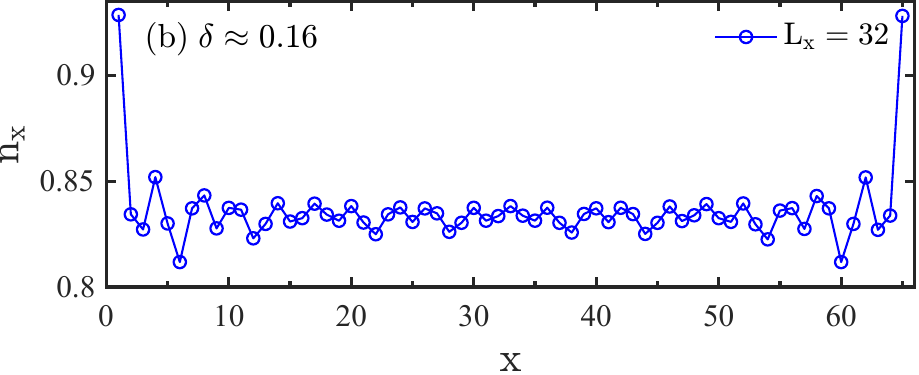} 
    \end{subfigure}
	\begin{subfigure}[b]{0.48\textwidth}
		\includegraphics[width=\textwidth]{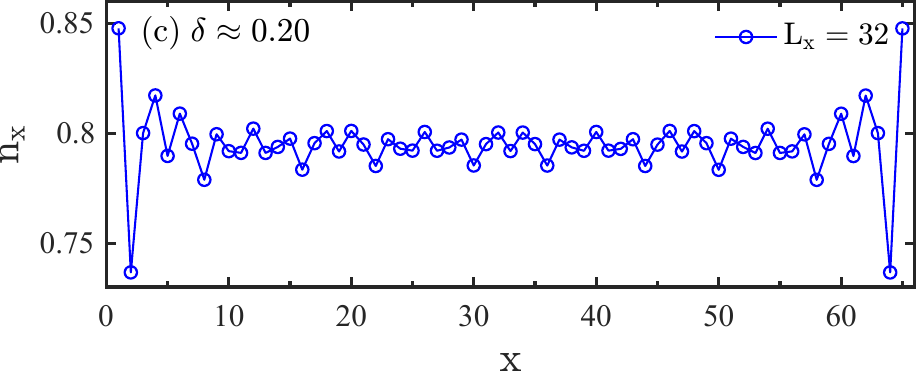} 
    \end{subfigure}
	\begin{subfigure}[b]{0.48\textwidth}
		\includegraphics[width=\textwidth]{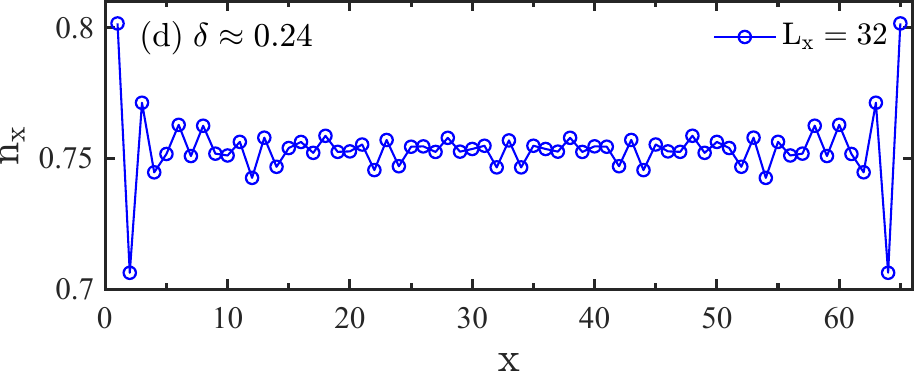} 
    \end{subfigure}
	\caption{\justifying Charge density profiles in the YC6 systems with $L_x=32$. $n(x,y)$ denotes the electron density at each lattice site, where $x$ ($y$) is the lattice index along the ${\bf e}_1$ (${\bf e}_2$) direction with the NN distance as the unit of length. The column-averaged electron density is denoted as $n_x$. (a) The hole density distribution $n_h(x,y) = 1 - n(x,y)$ for $\delta \approx 0.11$ in the CDW phase. (b)-(d) $n_x$ in the Fermi-liquid-like phase at $\delta \approx 0.16$, $0.20$, and $0.24$, respectively.}
    \label{charge}
\end{figure}

We first present the results of the charge density distributions on YC6 in Fig.~\ref{charge}.
The electron density at each lattice site $(x,y)$ is denoted as $n(x,y)$, where $x$ ($y$) is the lattice index along the direction ${\bf e}_1$ (${\bf e}_2$) with the NN distance as the unit of length. 
In the cases when electron density is uniform along the $\mathbf{e_2}$ direction, we employ the column-averaged electron density defined as $n_x \equiv \frac{1}{N_y} \sum_{y=1}^{N_y} n(x,y)$, where $N_y = 2L_y$ ($N_y = L_y$) for odd (even) $x$.

At $\delta \approx 0.11$, the CDW manifests itself as a Wigner crystal~\cite{kagome-tJ-Jiang-2017,kagome-tJlike-PCheng-2021}, as shown in Fig.~\ref{charge}(a). 
Here, we employ the hole distribution $n_h(x,y)=1-n(x,y)$ to highlight the visual effect.
The holes form a spot-like pattern with the density modulation amplitude $\approx 0.07$, indicating the formation of a larger unit cell. 
The number of doped holes matches exactly the number of yellow spots in Fig.~\ref{charge}(a), demonstrating that each emergent unit cell contains one hole on average. 
In the CDW phase, the emergent unit cells do not exhibit a fixed pattern but depend on the lattice geometry and doping ratio.
For example, on the YC6 cylinder, in addition to the spot-like pattern in Fig.~\ref{charge}(a), the hole density distributions form stripe-like patterns at $\delta = 1/36 - 1/18$~\cite{kagome-tJ-Jiang-2017}, with the number of emergent unit cells remaining equal to the number of doped holes. 

In the doping range $\delta \gtrsim 0.15$, the charge density distribution becomes uniform along the $\mathbf{e}_2$ direction, and therefore we analyze the column-averaged density $n(x)$ as shown in Figs.~\ref{charge}(b)-(d). 
Along the $\mathbf{e}_1$ direction, the charge density oscillation is significantly suppressed, which in the open boundary conditions strongly suggests the absence of a static CDW order. 
In Appendix~\ref{app:cdw}, we also show the distribution of the averaged charge density per unit cell.

\subsection{Correlation functions on YC6}
\begin{figure}[h] 
	\centering
	\begin{subfigure}[b]{0.48\textwidth}
		\includegraphics[width=0.494\textwidth]{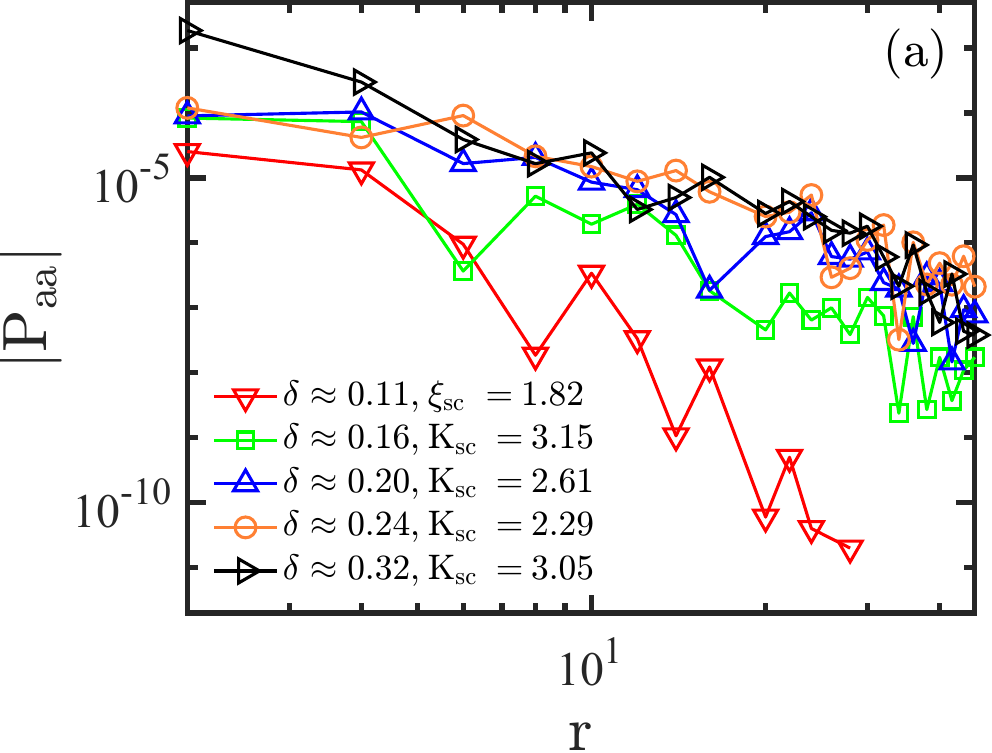} 
		\includegraphics[width=0.494\textwidth]{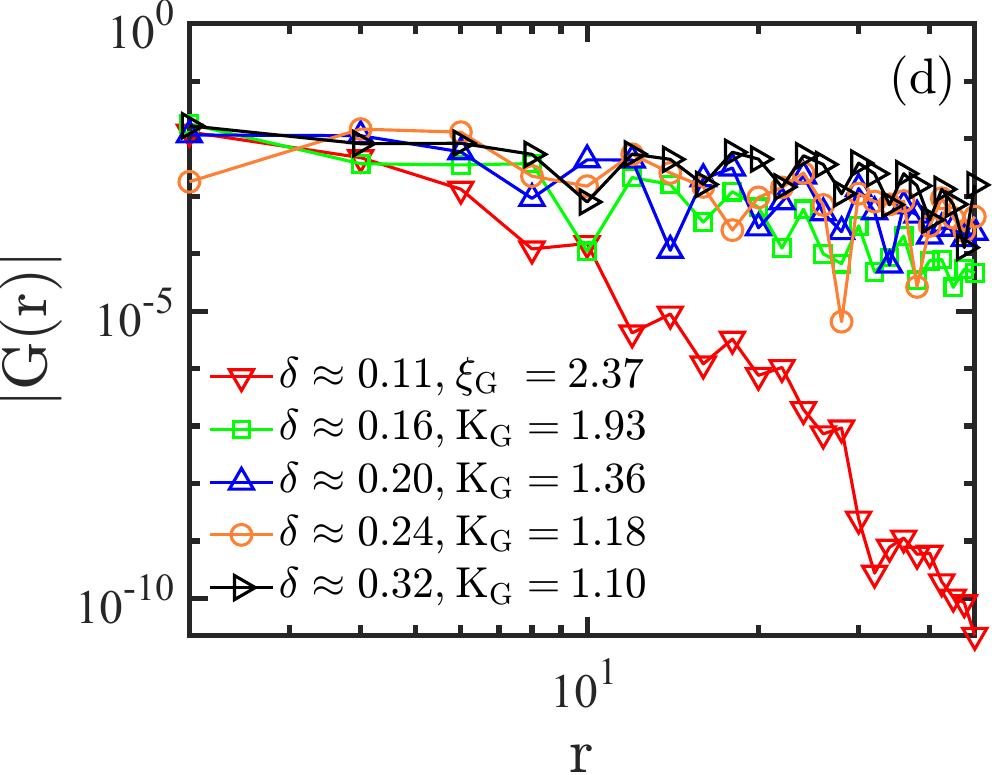} 
	\end{subfigure}
	\begin{subfigure}[b]{0.48\textwidth}
		\includegraphics[width=0.494\textwidth]{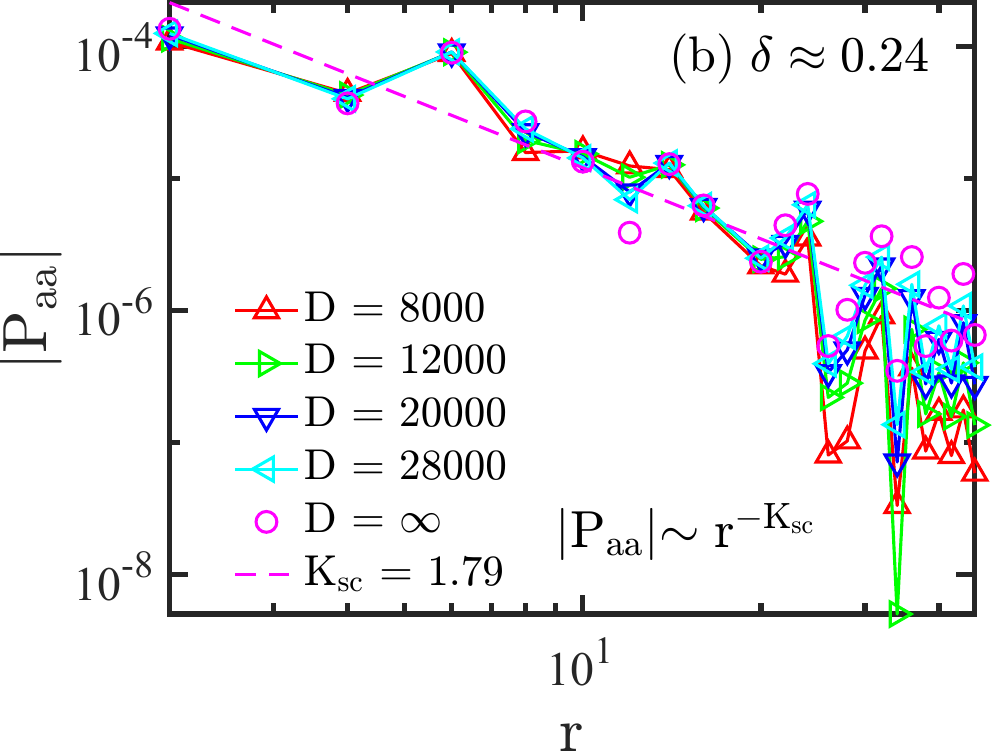} 
		\includegraphics[width=0.494\textwidth]{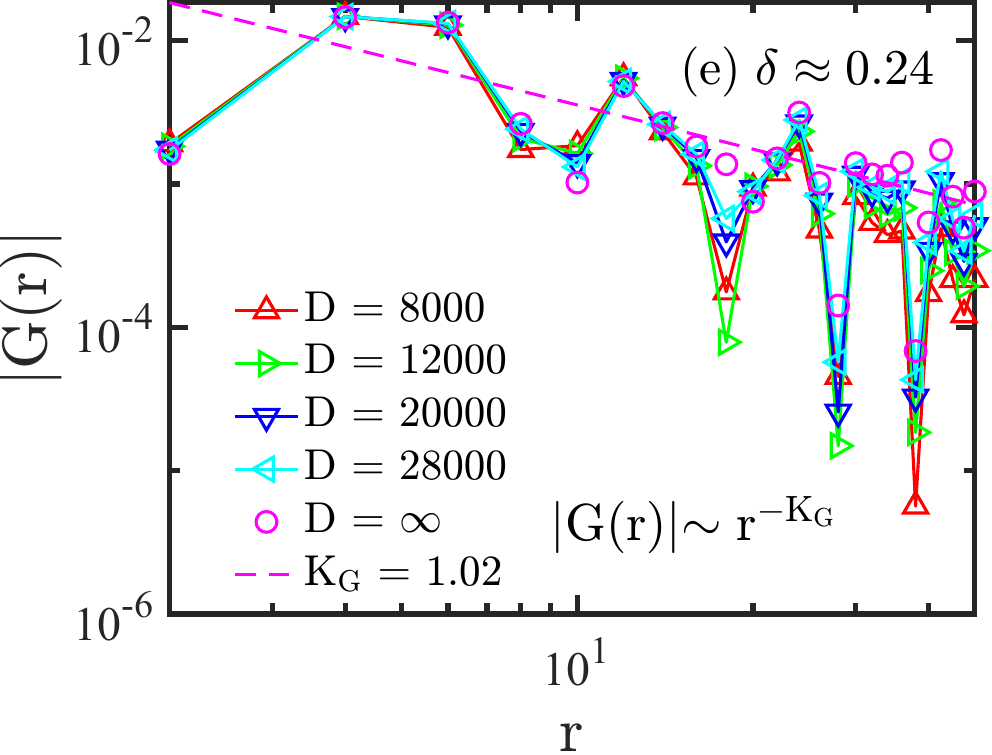} 
	\end{subfigure}
	\begin{subfigure}[b]{0.48\textwidth}
		\includegraphics[width=0.494\textwidth]{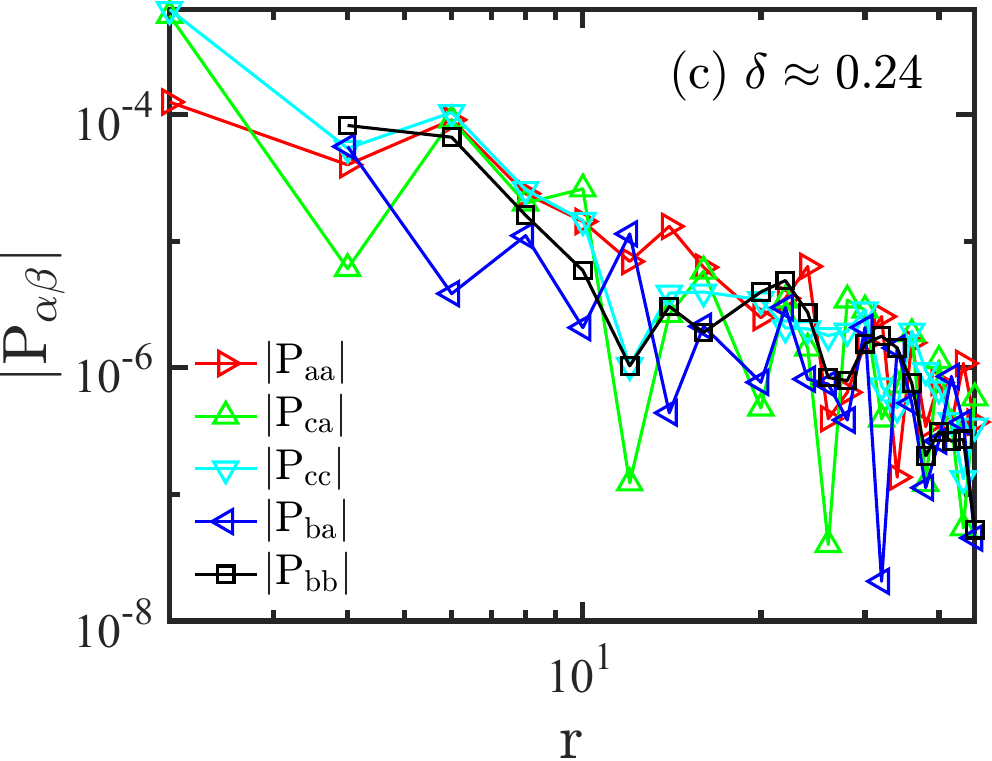} 
		\includegraphics[width=0.494\textwidth]{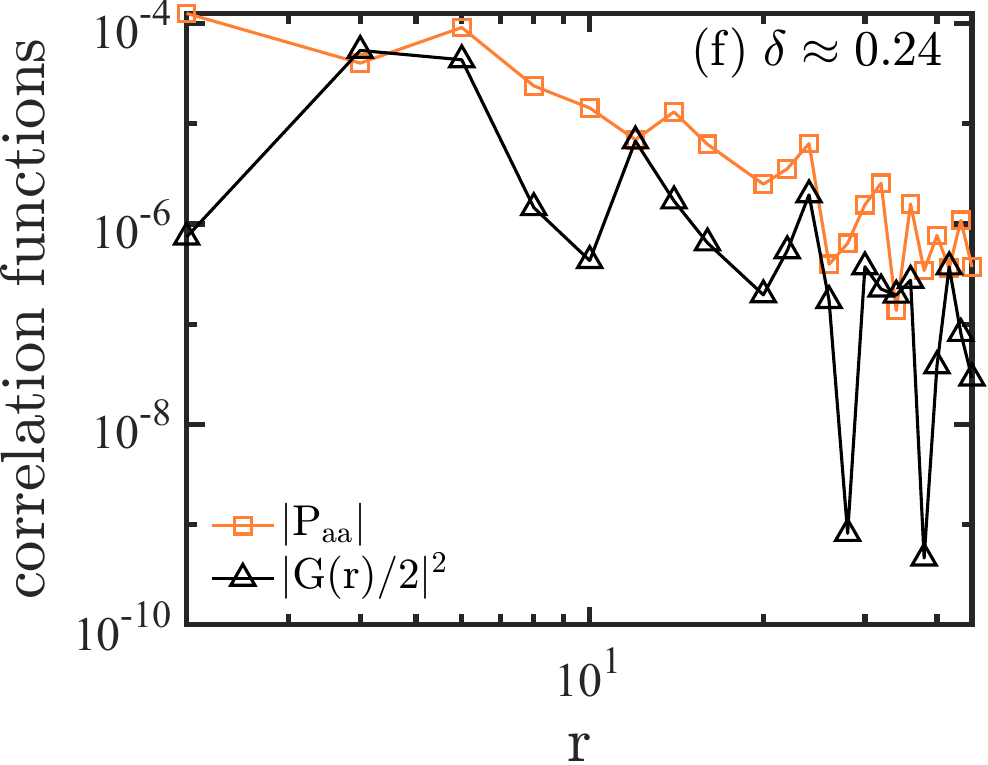} 
	\end{subfigure}
	\caption{\justifying SC pairing correlations and single-particle correlation $G(r)$ in the YC6 systems. (a) $|P_{aa}|$ for different doping levels. $\xi_{\rm sc}$ and $K_{\rm sc}$ are the fitted exponents in exponential and power-law decay, respectively. (b) $|P_{aa}|$ for $\delta \approx 0.24$ with different bond dimensions in the range of $D = 8000 - 28000$. The dashed line denotes the power-law fitting of the extrapolated $D \rightarrow \infty $ results. (c) Pairing correlation functions between different bonds $|P_{\alpha \beta}(r)|$ at $\delta \approx 0.24$. (d) and (e) are the similar plots for the single-particle Green’s function $G(r)$. (f) Comparison of pairing correlation $|P_{aa}|$ with the square of single-particle correlation $|G(r)/2|^2$ at $\delta \approx 0.24$.}
    \label{PrDr}
\end{figure}
In this subsection, we present the DMRG results of correlation functions. 
We begin by analyzing the spin-singlet pairing correlation function $P_{ \alpha\beta}(r)=\langle \hat{\Delta}_\alpha^\dagger(i_0) \hat{\Delta}_\beta(i_0 + r) \rangle$, where $i_0$ denotes the reference site at the $1/4$ length of the cylinder, and $r$ denotes the number of sites away from $i_0$ along the $\mathbf{e}_1$ direction. 
The spin-singlet pair-field creation operator $\hat{\Delta}_\alpha^\dagger(i)$ is defined as $\hat{\Delta}_\alpha^\dagger(i) = 1/\sqrt{2} ( \hat{c}_{i\uparrow}^\dagger \hat{c}_{i+\alpha\downarrow}^\dagger - \hat{c}_{i\downarrow}^\dagger \hat{c}_{i+\alpha\uparrow}^\dagger )$, where $\alpha$ denotes the three bond directions: $\mathbf{a}$, $\mathbf{c}$, and $\mathbf{b}$, which correspond to $\mathbf{e}_1/2$, $\mathbf{e}_2/2$, and $(\mathbf{e}_2 - \mathbf{e}_1)/2$, respectively [see Fig.~\ref{lattice_diagram}(a)]. 
Figure~\ref{PrDr}(a) shows $P_{aa}$ for $\delta \approx 0.11-0.32$, where $\xi_{\rm sc}$ and $K_{\rm sc}$ are the fitting exponents obtained in exponential decay function and power-law decay function, respectively.
From the CDW phase to the Fermi-liquid-like phase, the pairing correlation exhibits a substantial enhancement. 
For $\delta \approx 0.24$, we carefully analyze the results obtained by different bond dimensions (up to $D = 28000$), and we extrapolate the results to $D \rightarrow \infty$ limit using a polynomial function of bond dimension.
The extrapolated pairing correlation can be well fitted by a power-law decay with the power exponent $K_{\rm sc} \approx 1.79$ [Fig.~\ref{PrDr}(b)].
We have also examined other pairing correlation functions, including $P_{bb}$, $P_{cc}$, $P_{ca}$, and $P_{ba}$ [Fig.~\ref{PrDr}(c)], which have very close magnitudes and power exponents. 
For simplicity, we demonstrate the results of $P_{aa}$ as a representative.

Next, we examine the single-particle Green's function $G(r) = \sum_\sigma\langle \hat{c}_{i_0,\sigma}^\dagger \hat{c}_{i_0+r,\sigma}\rangle$. 
In the CDW phase, $G(r)$ decays exponentially [Fig.~\ref{PrDr}(d)], characterizing a charge insulating phase.
In the Fermi-liquid-like phase, $G(r)$ increases significantly and decays algebraically. 
Similarly, we extract the $D \rightarrow \infty$ results. 
For $\delta \approx 0.24$, the extrapolated $G(r)$ fits the power-law decay quite well with the power exponent $K_{\rm G} \gtrsim 1$ [see Fig.~\ref{PrDr}(e)], showing the gapless single-partle excitations. 
To further clarify the nature of pairing, we compare $P_{aa}$ with the square of single-particle correlation $(G(r)/2)^2$ and find that the two quantities are very close [Fig.~\ref{PrDr}(f)], showing the absence of hole pairing in this phase.

\begin{figure}[h] 
	\centering
	\begin{subfigure}[b]{0.48\textwidth}
		\includegraphics[width=0.494\textwidth]{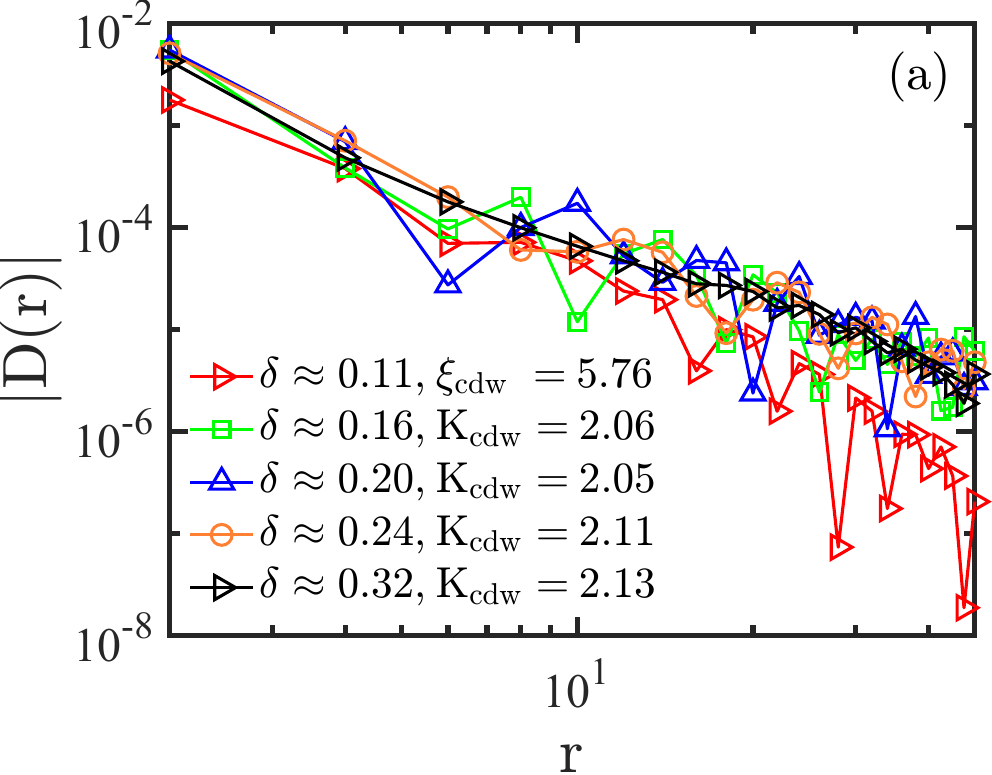} 
		\includegraphics[width=0.494\textwidth]{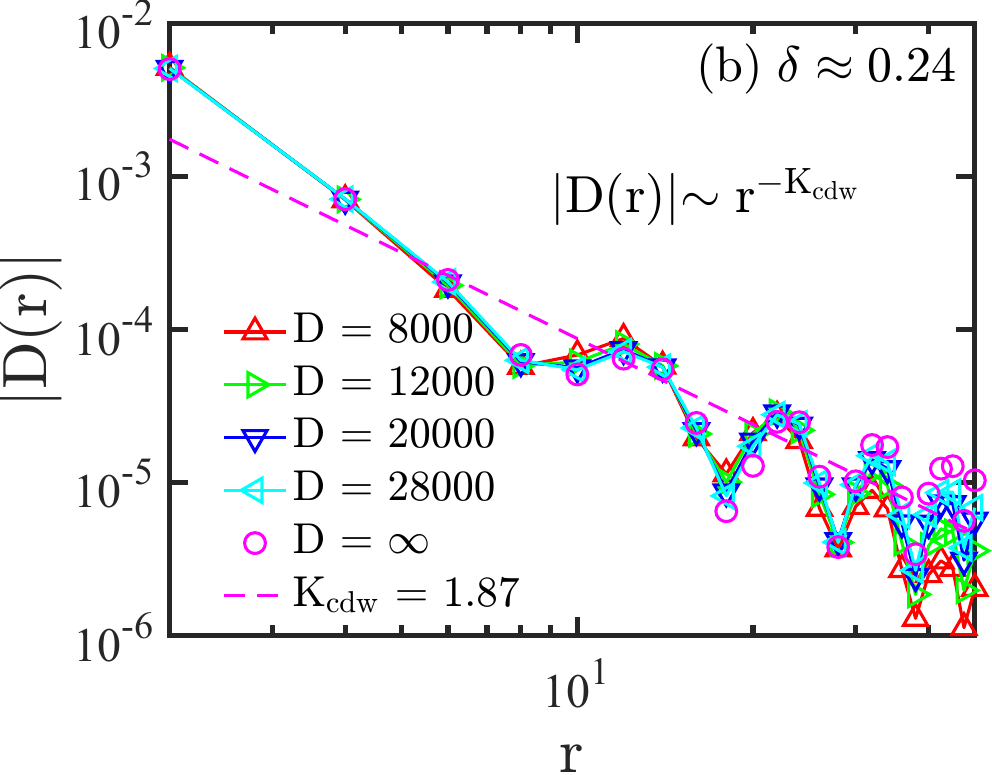} 
	\end{subfigure}
	\begin{subfigure}[b]{0.48\textwidth}
		\includegraphics[width=0.494\textwidth]{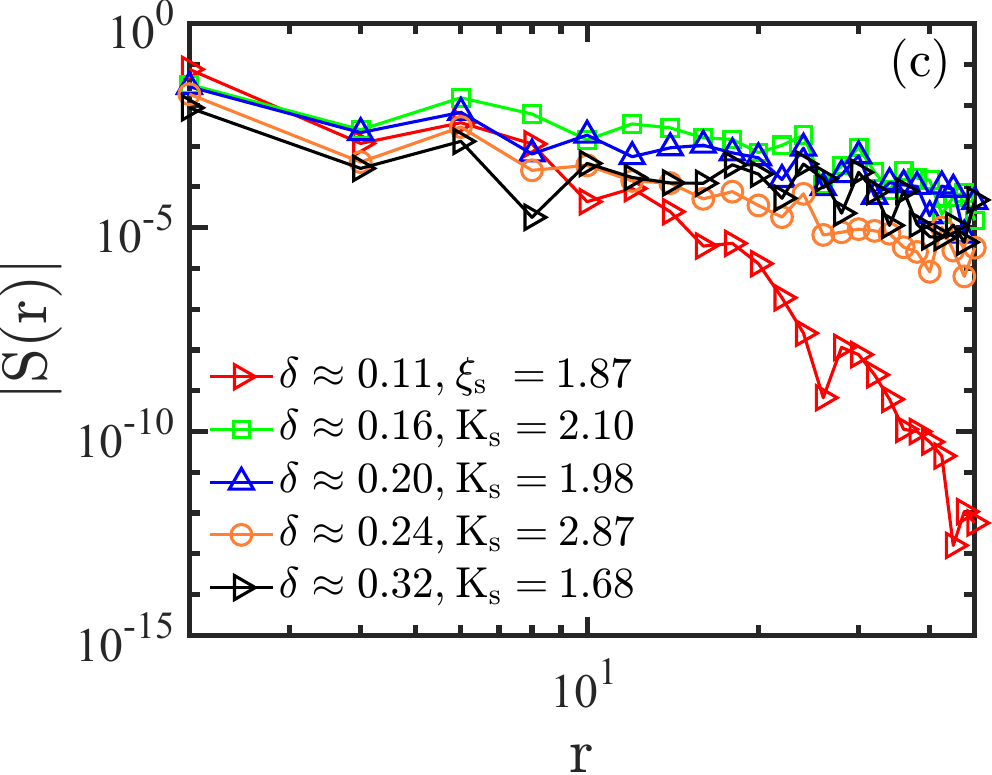} 
		\includegraphics[width=0.494\textwidth]{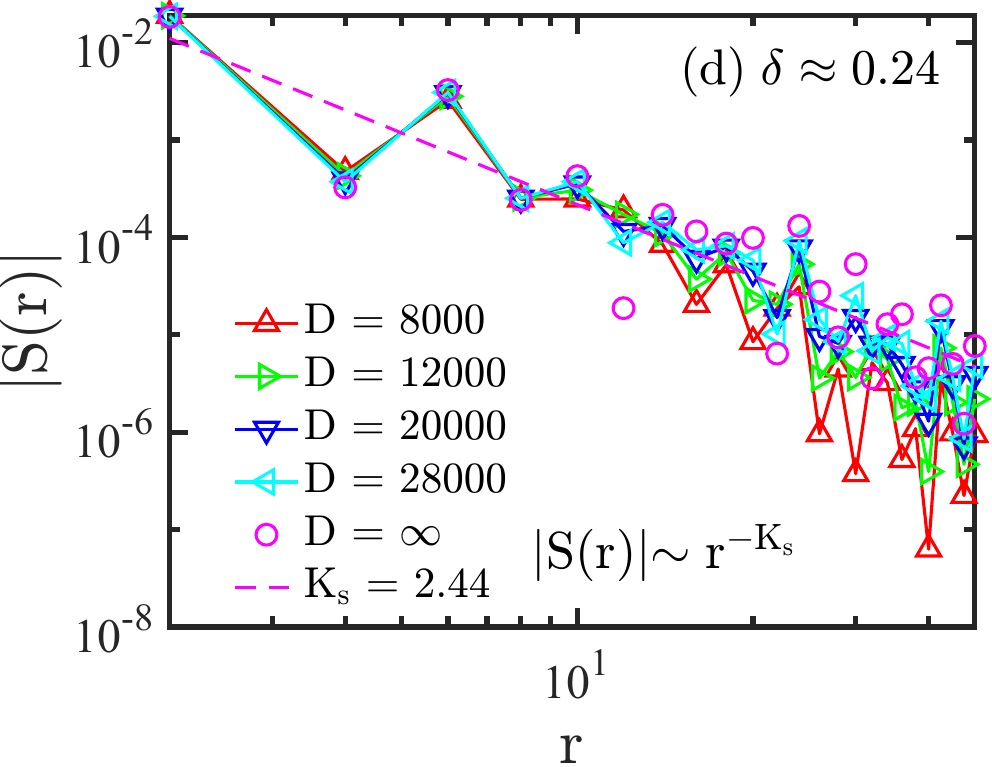} 
	\end{subfigure}
    \begin{subfigure}[b]{0.48\textwidth}
		\includegraphics[width=0.494\textwidth]{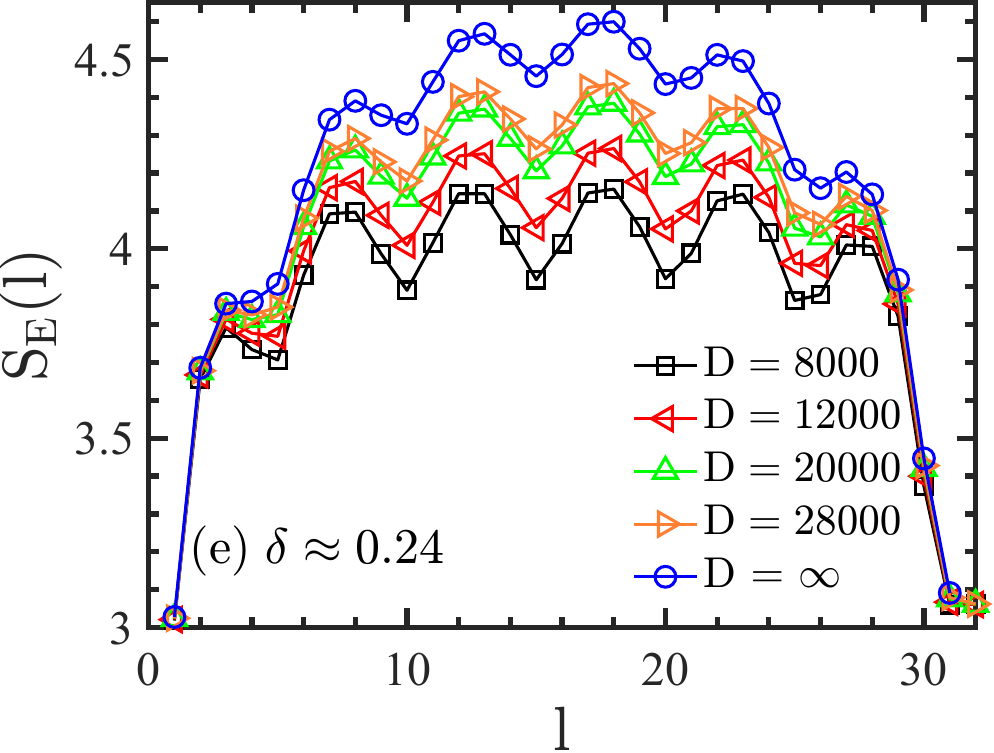} 
		\includegraphics[width=0.494\textwidth]{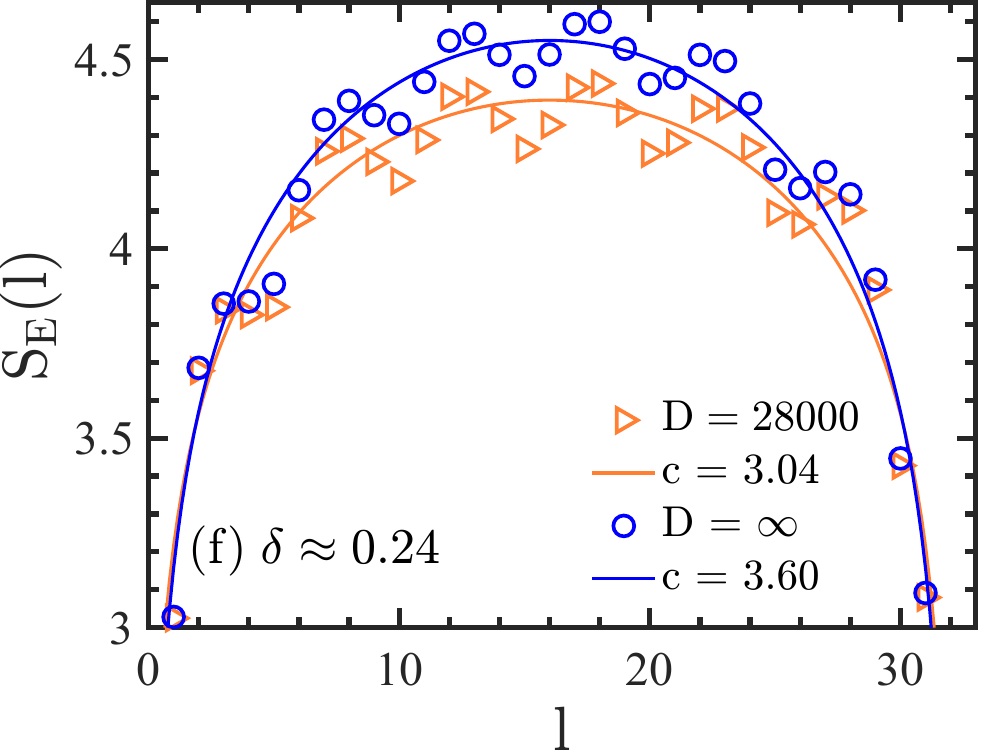} 
	\end{subfigure}
	\caption{\justifying Density correlation $D(r)$, spin correlation $S(r)$, and entanglement entropy $S_E$ in the YC6 systems with $L_x=32$. (a) $|D(r)|$ for different doping levels. (b) $|D(r)|$ for $\delta \approx 0.24$ at different bond dimensions $D = 8000-28000$. (c) and (d) are the similar plots for the spin correlation $S(r)$. (e) Entanglement entropy for $\delta \approx 0.24$ at different bond dimensions. The $D = \infty$ results are extrapolated from the finite bond-dimension data. The label $l$ denotes the unit cell index along the ${\bf e}_1$ direction. (f) Fitting of the central charge $c$ from the formula $S_E(l) = (c/6)\ln \left[(L_x/\pi) \sin \left(l \pi/L_x \right) \right] + g$ for the $D = 28000$ and $D = \infty$ data, giving a finite central charge. }
    \label{DrSr}
\end{figure}

We then show the results of the density correlation function $D(r)=\langle \hat{n}_{i_0} \hat{n}_{i_0+r} \rangle - \langle \hat{n}_{i_0} \rangle \langle \hat{n}_{i_0+r} \rangle$. 
At $\delta \approx 0.11$, the charge density profile shows a Wigner crystal feature with spontaneous translational symmetry breaking in both ${\bf e}_1$ and ${\bf e}_2$ directions [Fig.~\ref{charge}(a)].
In this state, our defined density correlation function may describe the charge density fluctuations. 
In the Fermi-liquid-like phase, the static CDW order disappears [Figs.~\ref{charge}(b)-(d)], and this density correlation function can describe the charge order.
As shown in Fig.~\ref{DrSr}(a), the decay of $D(r)$ can be fitted algebraically quite well with the power exponent $K_{\rm cdw} \approx 2$, which is consistent with a Fermi-liquid-like state. 
The analysis of the bond dimension dependence shows the very good convergence of density correlation function [Fig.~\ref{DrSr}(b)].

For the spin correlation function $S(r)=\langle {\bf S}_{i_0} \cdot {\bf S}_{i_0+r}\rangle$, it shows an exponential decay with a short correlation length in the CDW phase [Fig.~\ref{DrSr}(c) and Ref.~\cite{kagome-tJ-Jiang-2017}].
In the Fermi-liquid-like phase, spin correlation is enhanced significantly and appears to decay algebraically.
Through the bond-dimension extrapolation, the well-converged spin correlation function can fit a power-law decay, with the exponent $K_{\rm s} \approx 2.4$ for $\delta \approx 0.24$ [Fig.~\ref{DrSr}(d)].
These correlation function results all agree with a Fermi-liquid-like state.

\subsection{Entanglement entropy on YC6}
In this subsection, we calculate the entanglement entropy and fit the central charge to characterize the gapless nature of the Fermi-liquid-like phase. 
The entanglement entropy between two subsystems is defined as $S_E(l) = - \text{Tr}[\rho(l) \ln \rho(l)]$ ($1\leq l \leq L_x-1$), where $l$ and $L_x - l$ are the lengths of the two subsystems, and $\rho(l)$ is the reduced density matrix of the subsystem $l$ obtained from the ground state of the whole system.
According to the conformal field theory, for one-dimensional critical systems with open boundary conditions, the bipartite entanglement entropy $S_E$ follows the formula $S_E(l) = (c/6)\ln \left[(L_x/\pi) \sin \left(l \pi/L_x \right) \right] + g$, where $c$ is the central charge, $L_x$ is the length of the whole system and $g$ is a non-universal constant~\cite{Entanglement_entropy_2004}. 
In Fig.~\ref{DrSr}(e), we show the entanglement entropy $S_E$ for $\delta \approx 0.24$ on the $L_x=32$ cylinder. 
With increasing bond dimensions, $S_E$ continues to grow. 
For the entropy at each given $l$, we extrapolate the $D \rightarrow \infty$ result from the DMRG data obtained by different bond dimensions (see Appendix~\ref{app:entropy}).
In Fig.~\ref{DrSr}(f), we fit the entropy data to extract the central charge.
For $D = 28000$, we find $c \approx 3$, and for $D \rightarrow \infty$ we obtain $c \approx 3.6$.
Notice that to determine the central charge, one also needs to enlarge the system length and extract the central charge in the $L_x \rightarrow \infty$ limit.
Limited by computational cost, we may not accurately determine the value of the central charge, but its finite value clearly characterizes the gapless nature of this state.

\subsection{DMRG results on YC8}
\begin{figure}[h] 
	\centering
	\begin{subfigure}[b]{0.48\textwidth}
		\includegraphics[width=0.494\textwidth]{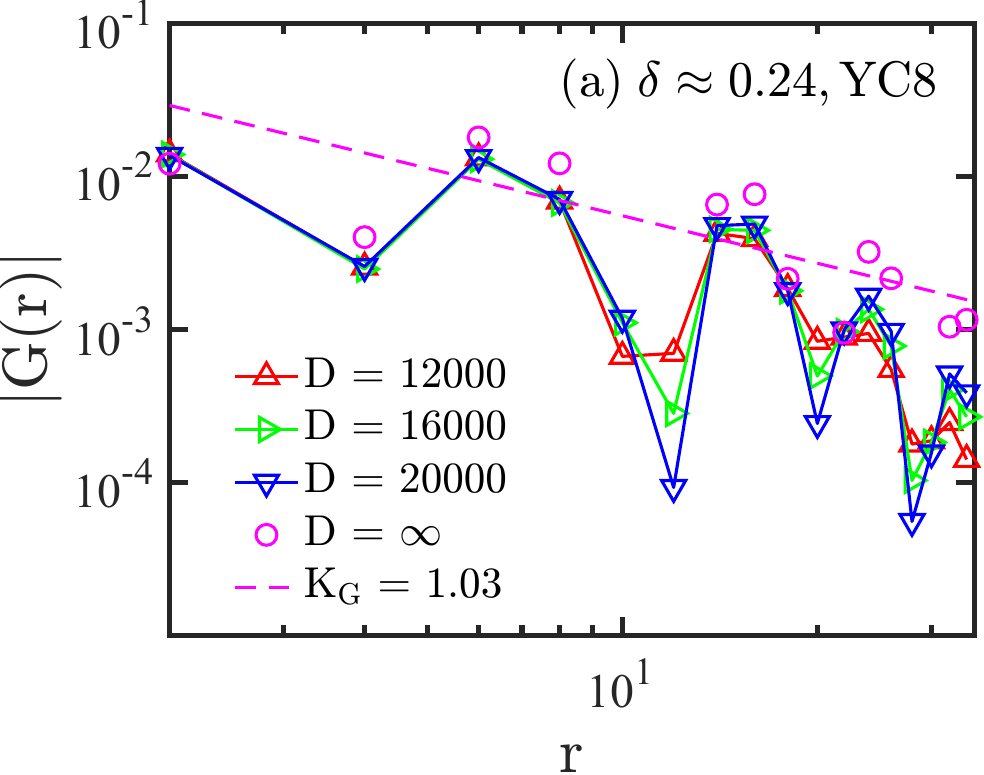} 
		\includegraphics[width=0.494\textwidth]{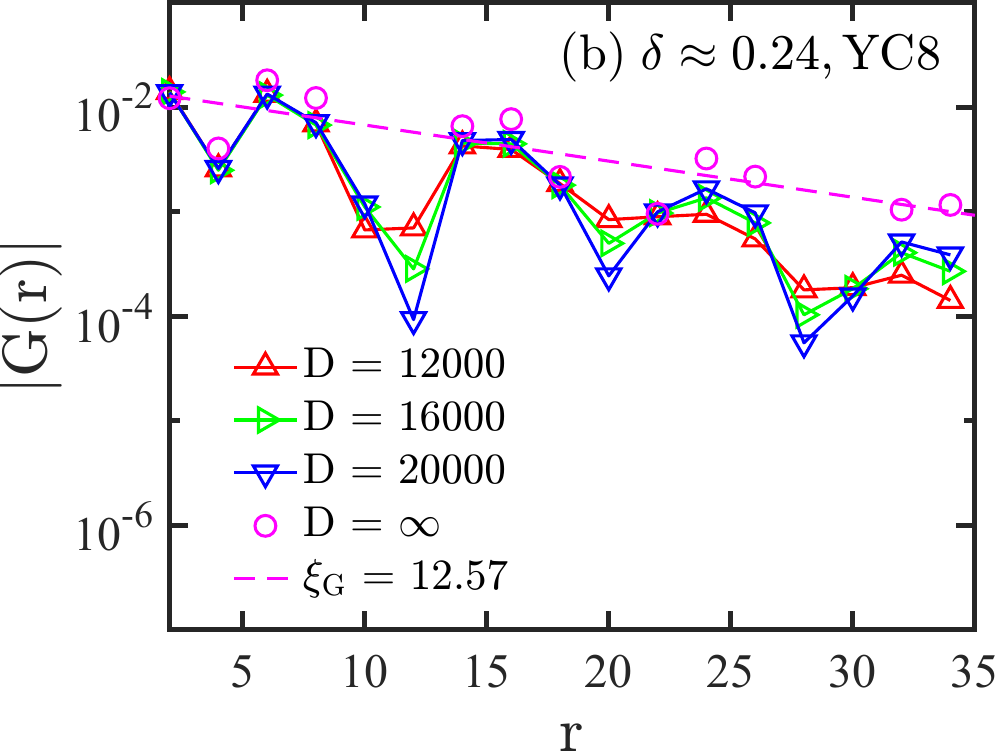} 
	\end{subfigure}
    \begin{subfigure}[b]{0.48\textwidth}
		\includegraphics[width=0.494\textwidth]{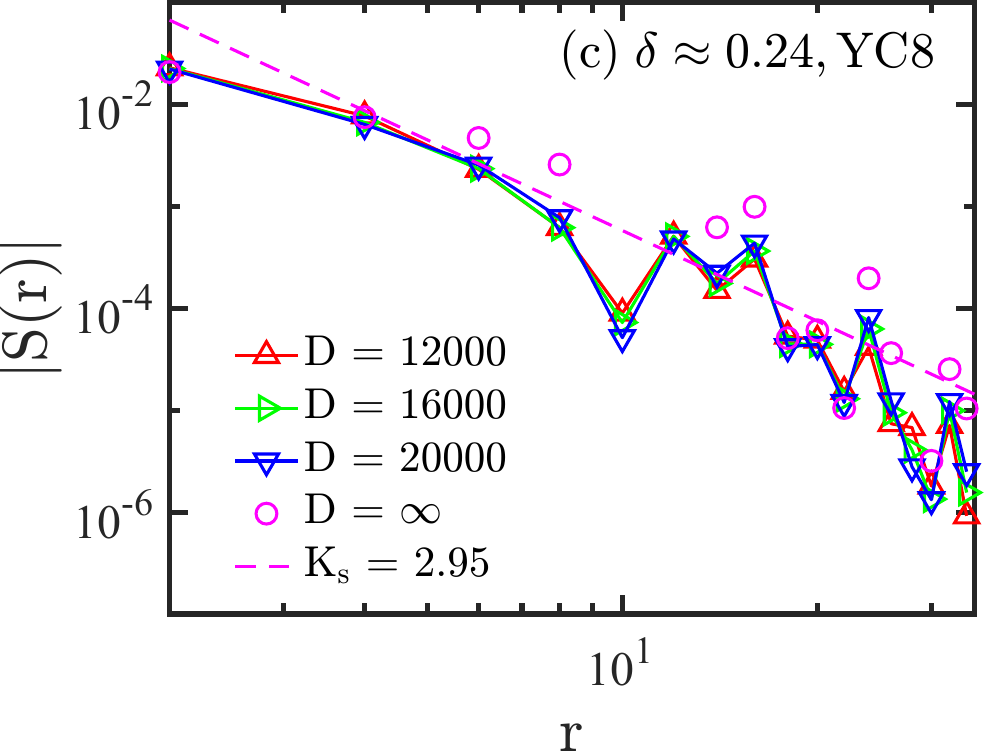} 
		\includegraphics[width=0.494\textwidth]{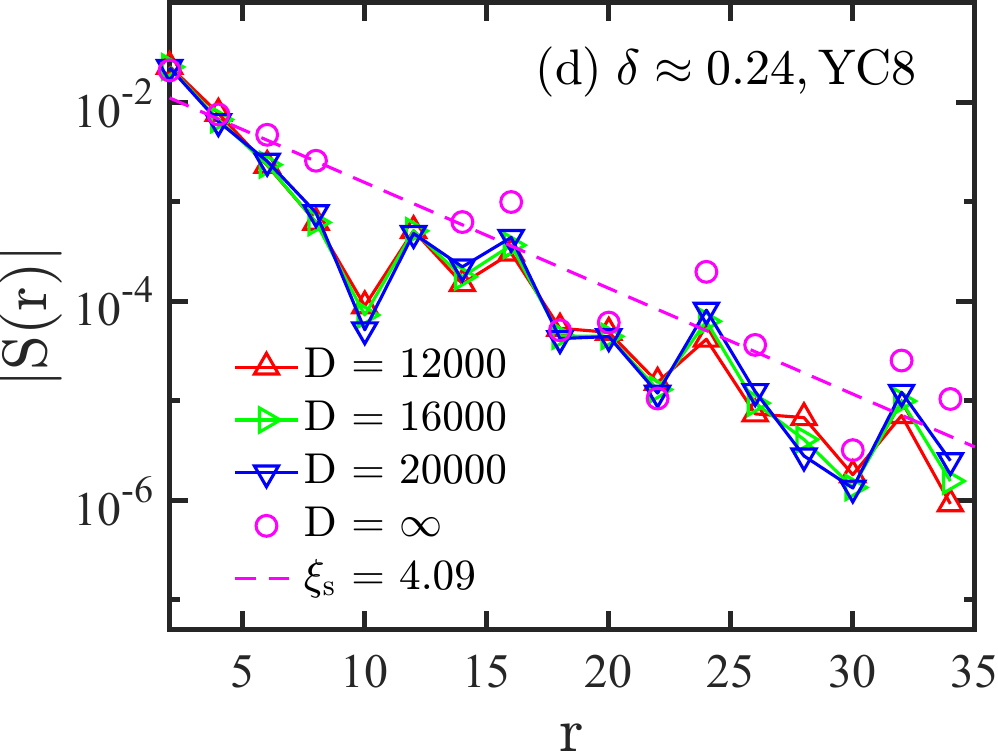}
	\end{subfigure}
	\begin{subfigure}[b]{0.48\textwidth}
		\includegraphics[width=0.494\textwidth]{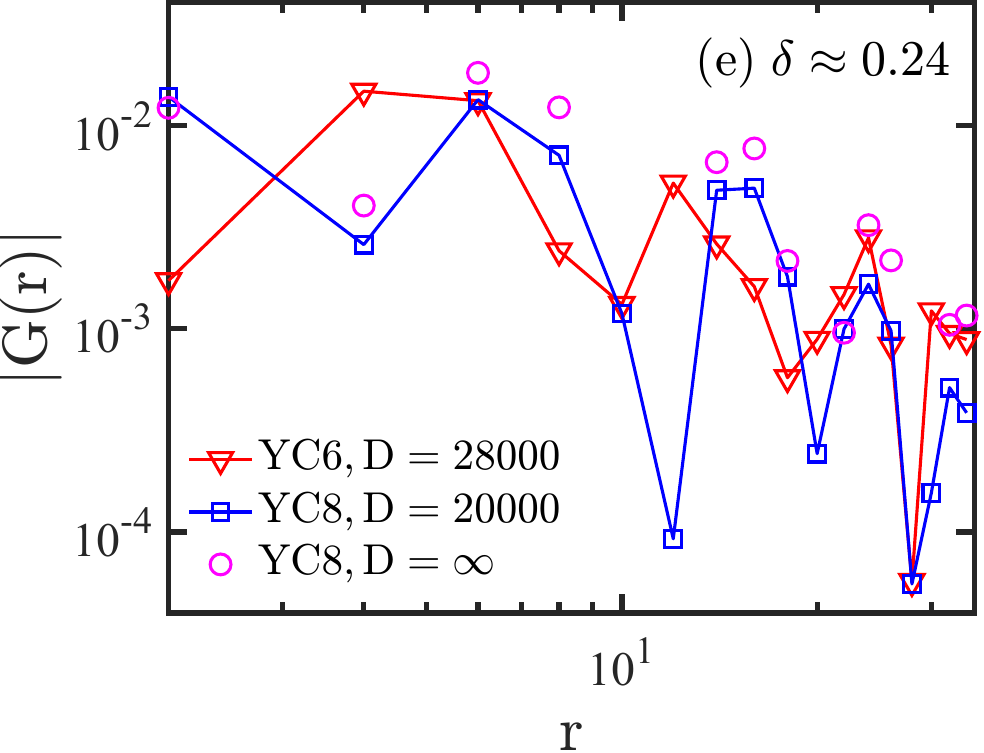} 
		\includegraphics[width=0.494\textwidth]{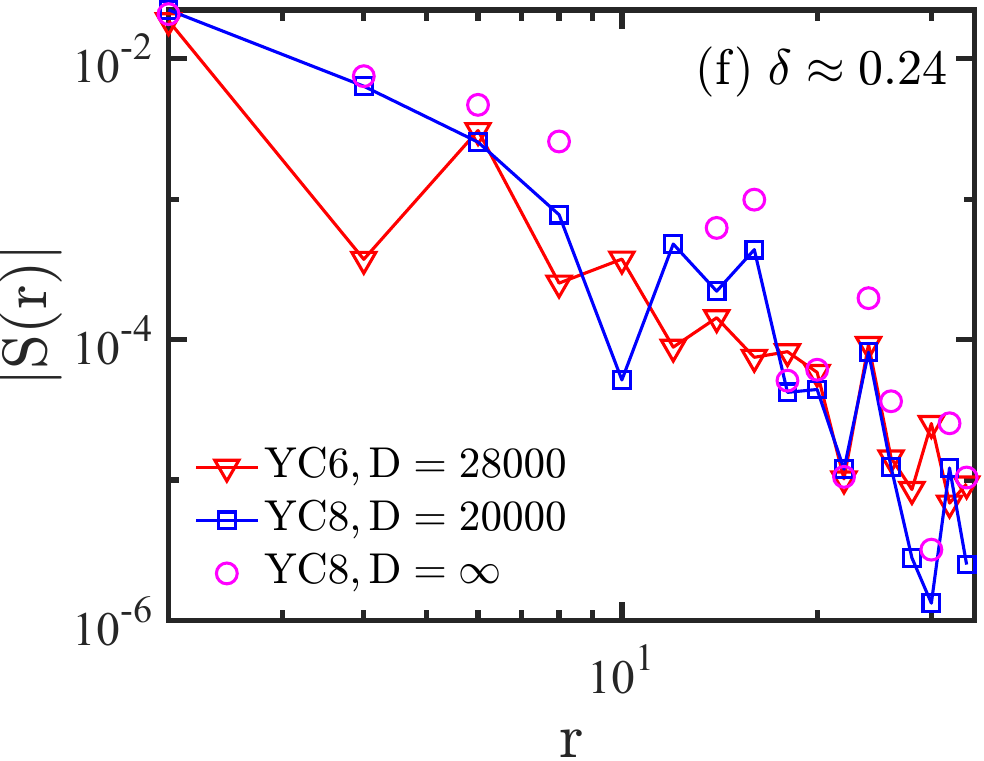} 
	\end{subfigure}
    \begin{subfigure}[b]{0.48\textwidth}
		\includegraphics[width=0.494\textwidth]{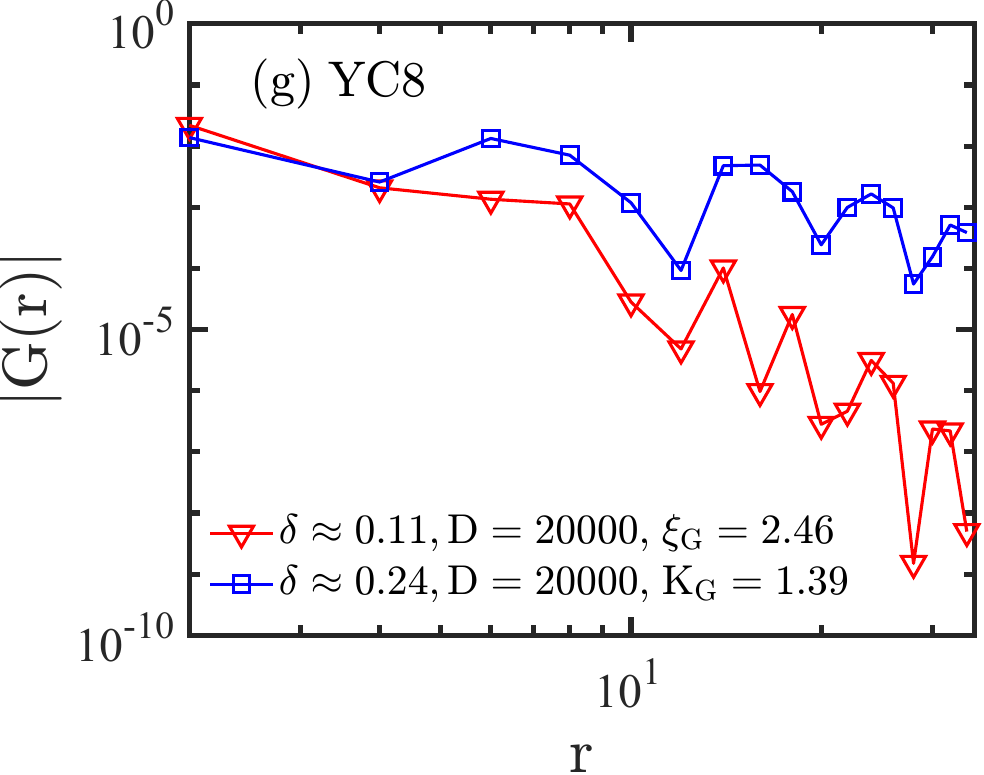} 
		\includegraphics[width=0.494\textwidth]{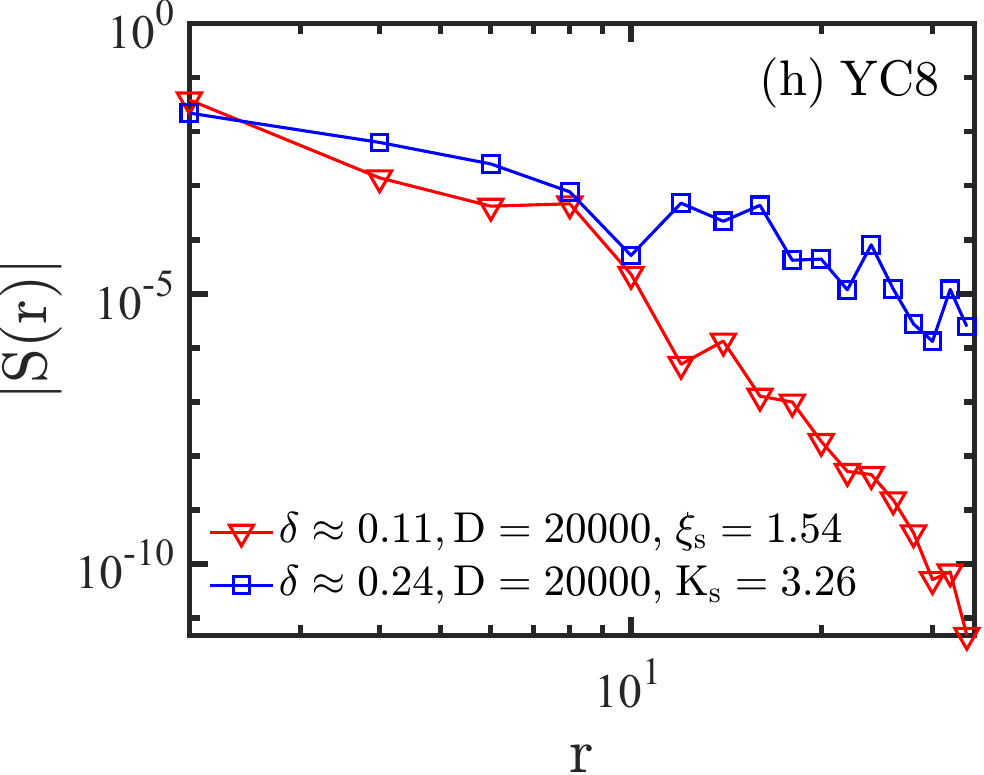} 
	\end{subfigure}
	\caption{\justifying Single-particle Green's function $G(r)$ and spin correlation $S(r)$ in the YC8 systems. (a) Double-logarithmic plot and (b) semi-logarithmic plot of $G(r)$ for $\delta\approx0.24$ on the YC8 cylinder, with the bond dimensions $D = 12000-20000$. (c) and (d) Similar plots of spin correlation $S(r)$. (e) and (f) Comparisons of $G(r)$ and $S(r)$ between YC6 and YC8 cylinders at $\delta \approx 0.24$. (g) and (h) Comparisons of $G(r)$ and $S(r)$ at $\delta \approx 0.11$ and $\delta \approx 0.24$ on the YC8 cylinder.}
    \label{YC8_Cr}
\end{figure}
We further study the wider YC8 system to examine finite-size effects and gain insight into the ground state in the two-dimensional limit.
In Figs.~\ref{YC8_Cr}(a)-(d), we present the single-particle Green's function $G(r)$ and spin correlation $S(r)$ of the YC8 system at $\delta \approx 0.24$, which are obtained with the bond dimensions $D = 12000 - 20000$.
Here, the bond dimensions are smaller than $D = 28000$ on YC6 due to the computational cost.
We extrapolate the data with respect to the bond dimension to estimate the results in the $D \to \infty$ limit. 
In this extrapolation, we discard the data points that oscillate with the bond dimension. 
To analyze the decay behavior of the correlation functions, we consider both double–logarithmic and semi–logarithmic fitting.
For $G(r)$, the power-law fitting up to the distance $r \sim 30$ gives an exponent $K_{\rm G} \approx 1.03$ [Fig.~\ref{YC8_Cr}(a)], and the exponential fitting gives a correlation length $\xi_{\rm G} \approx 12.6$ [Fig.~\ref{YC8_Cr}(b)], indicating a slow decay of $G(r)$.
For $S(r)$, the power-law fitting gives $K_{\rm s} \approx 2.95$ [Fig.~\ref{YC8_Cr}(c)] and the exponential fitting obtains $\xi_{\rm s} \approx 4.1$ [Fig.~\ref{YC8_Cr}(d)].

To analyze the size dependence, we further compare the results of the YC6 and YC8 systems at $\delta \approx 0.24$, for both $G(r)$ [Fig.~\ref{YC8_Cr}(e)] and $S(r)$ [Fig.~\ref{YC8_Cr}(f)]. 
Overall, the results of the two systems are consistent, and only the longer distance correlations ($r \gtrsim 25$) of the YC8 system display a more pronounced trend of decrease, which however may be due to the less convergence.
By keeping larger bond dimensions, one can expect an enhancement of the long-distance data.
As an estimate, we also show the extrapolated $D \rightarrow \infty$ results of YC8, which are indeed more consistent with the YC6 results.

To distinguish the CDW and Fermi-liquid-like states on YC8, we also compare the correlation functions at $\delta \approx 0.11$ and $\delta \approx 0.24$, as shown in Figs.~\ref{YC8_Cr}(g) and (h). 
Both $G(r)$ and $S(r)$ decay exponentially with short correlation lengths at $\delta \approx 0.11$, but are enhanced by several orders of magnitude at $\delta \approx 0.24$. 
These features are consistent with our observations on the YC6 cylinder.
Although the YC8 results may be less convergent, the systematic comparisons in Fig.~\ref{YC8_Cr} also suggest a stable Fermi-liquid-like state on larger system size.

\section{The quantum states near $1/3$ doping}
\begin{figure}[h] 
	\centering
	\begin{subfigure}[b]{0.48\textwidth}
		\includegraphics[width=\textwidth]{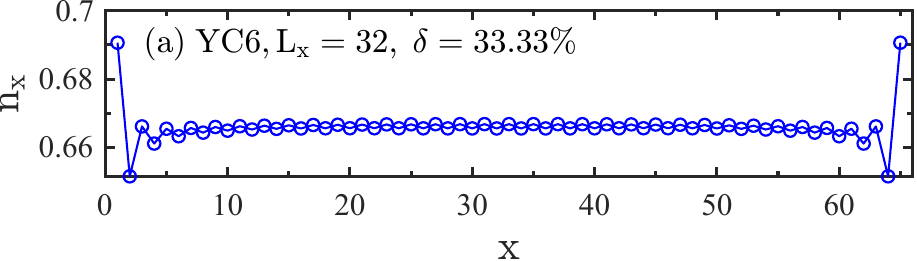} 
	\end{subfigure}
    \begin{subfigure}[b]{0.48\textwidth}
		\includegraphics[width=0.494\textwidth]{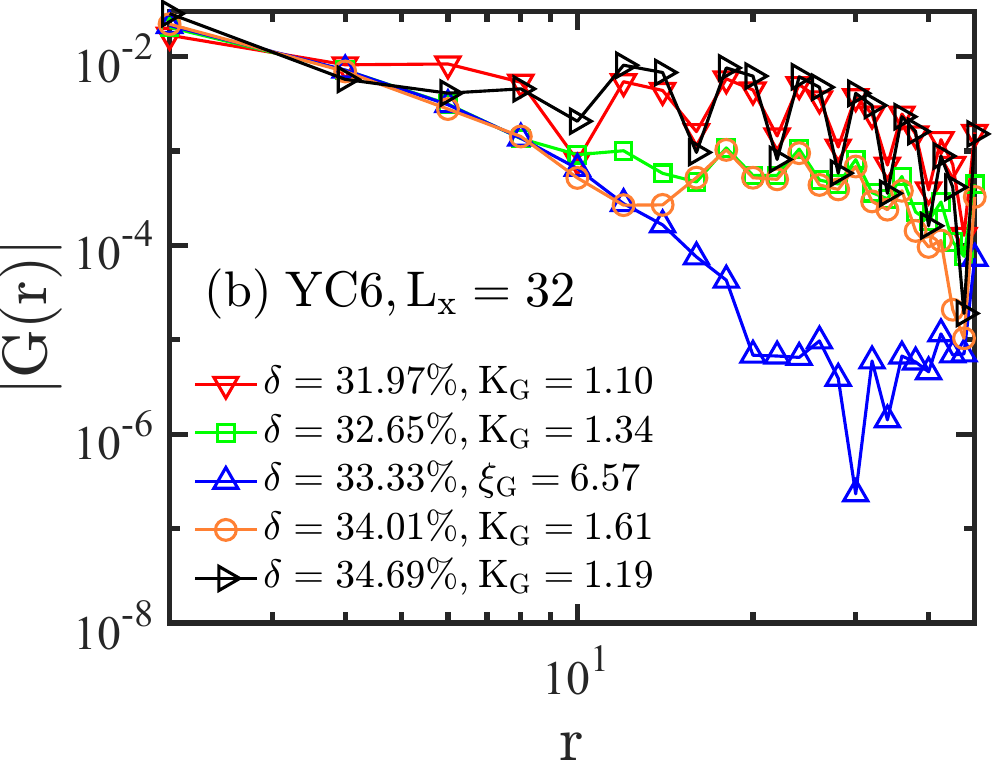} 
		\includegraphics[width=0.494\textwidth]{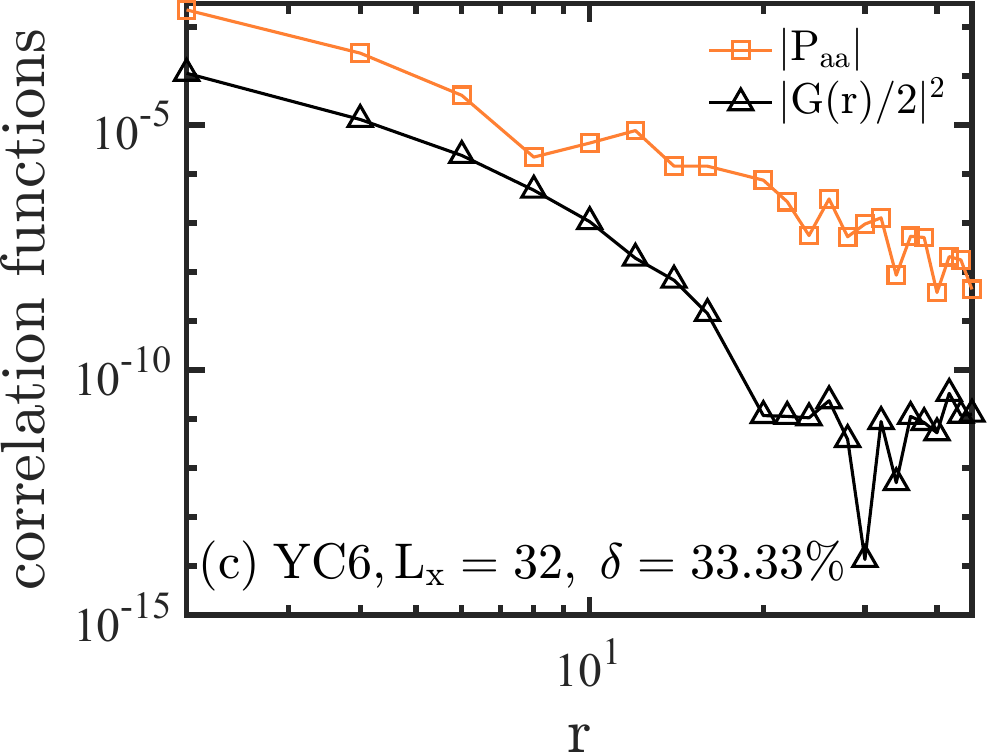}
	\end{subfigure}
	\begin{subfigure}[b]{0.48\textwidth}
		\includegraphics[width=0.494\textwidth]{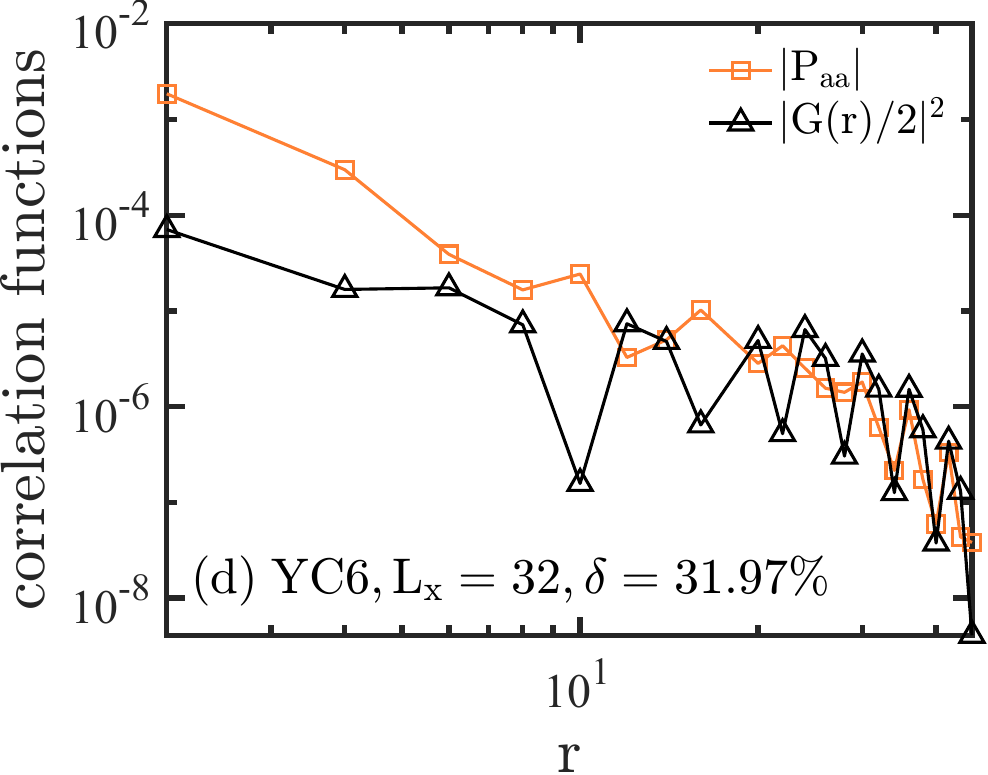} 
		\includegraphics[width=0.494\textwidth]{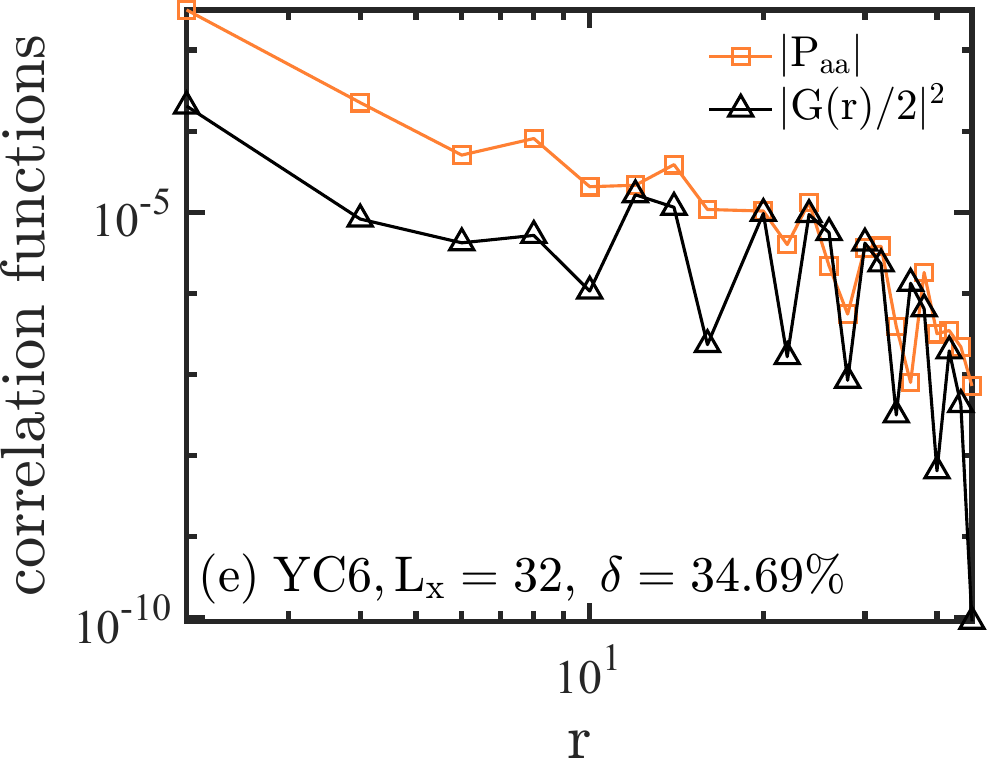} 
	\end{subfigure}
    \begin{subfigure}[b]{0.48\textwidth}
		\includegraphics[width=0.494\textwidth]{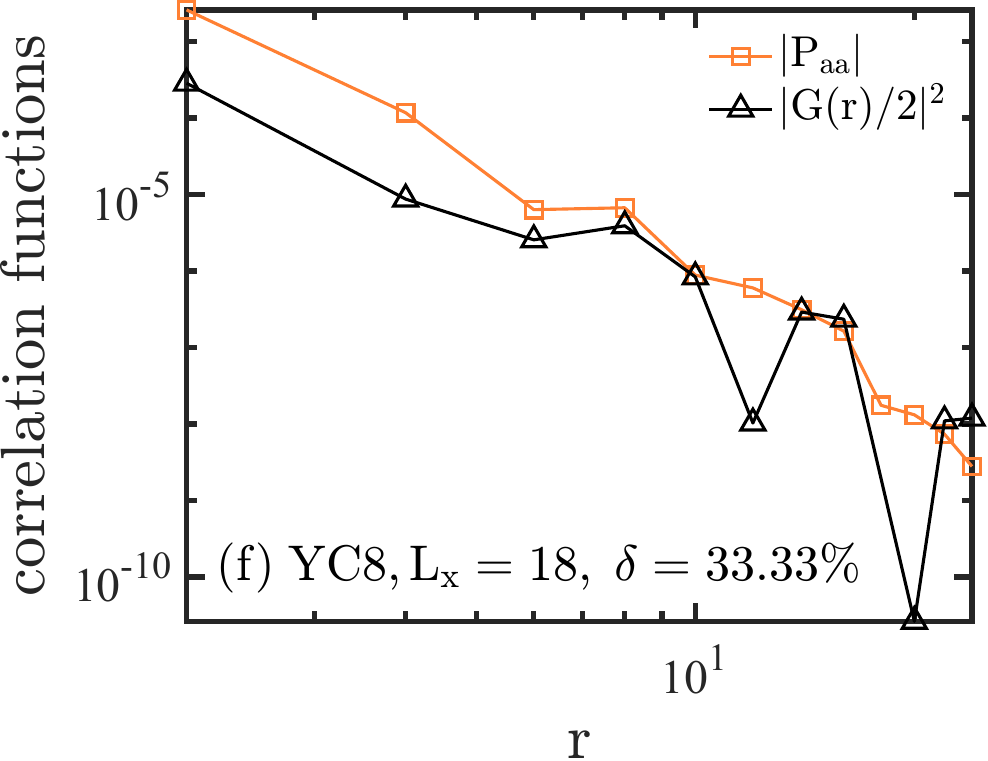} 
		\includegraphics[width=0.494\textwidth]{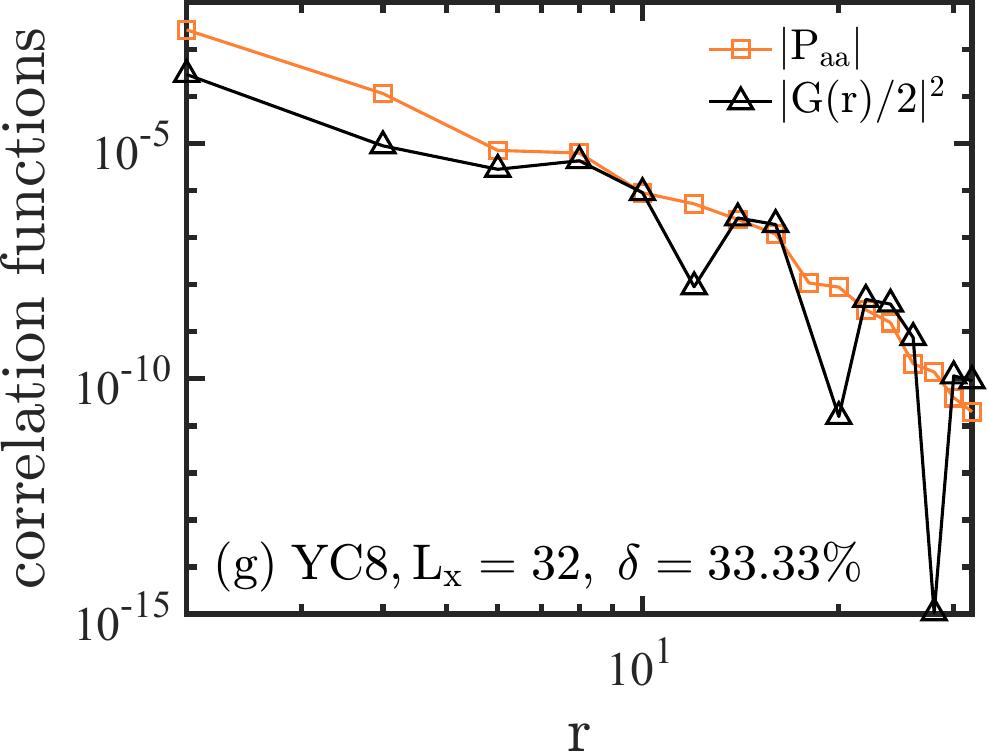} 
	\end{subfigure}
	\caption{\justifying Charge density profiles and correlation functions around $1/3$ doping. (a) Charge density profiles at $\delta = 1/3$. (b) Single-particle Green’s function $G(r)$ for several doping levels around $\delta = 1/3$. (c)-(e) Comparisons between the pairing correlation $|P_{aa}|$ and the square of single-particle Green’s function $|G(r)/2|^2$, at $\delta = 1/3$, $31.97\%$, and $34.69\%$, respectively. The quantities in (a)-(e) are measured on the YC6 cylinder with the bond dimensions $D = 16000$. (f) and (g) Comparisons between the pairing correlation $|P_{aa}|$ and the square of single-particle Green’s function $|G(r)/2|^2$ at $\delta = 1/3$ on the YC8 cylinder with $L_x=18$ ($D = 24000$) and $L_x=32$ ($D = 16000$), respectively. It should be noted that in the calculations of (f) and (g), we used the regular cylinder geometry without the additional right boundary, which allows the precise $1/3$ doping on the YC8 cylinder.}
    \label{033}
\end{figure}

In this section, we focus on the DMRG results near $1/3$ doping. 
Since our DMRG simulations study the ground state in the sector with total spin quantum number $S=0$, the number of holes can only be changed by even numbers.
Here, we investigate the doping ratio $\delta = 1/3$ ($33.33\%$) and the nearby doping levels that differ by two holes ($32.65\%$ and $34.01\%$) and by four holes ($31.97\%$ and $34.69\%$).
The results are shown in Fig.~\ref{033}.

On YC6, the charge density profile is nearly uniform at $\delta = 1/3$ [Fig.~\ref{033}(a)], showing the absence of a static CDW order.
The doping dependence of single-particle correlation [Fig.~\ref{033}(b)] shows that $|G(r)|$ are suppressed at $\delta = 32.65\%$, $33.33\%$, and $34.01\%$.
In particular, $|G(r)|$ exhibits a pronounced drop at $\delta = 1/3$, which behaves like an exponential decay.
At the higher $\delta = 34.69\%$, $|G(r)|$ quickly recovers the magnitudes in the Fermi-liquid-like phase. 
This observation suggests a new state in a narrow doping range near $\delta = 1/3$, which appears to open the single-particle gap.

To further clarify the nature of this state, we compare the square of single-particle correlation with pairing correlation.
As shown in Fig.~\ref{033}(c), the pairing correlation is much stronger than $|G(r)/2|^2$ at $\delta=1/3$, which suggests the formed hole pairing. 
At $\delta = 31.97\%$ and $34.69\%$, the two quantities show comparable magnitudes as illustrated in Figs.~\ref{033}(d)-(e), which agree with the expected feature in the Fermi-liquid-like phase. 
In addition, we have also compared the spin correlation function at $\delta = 1/3$ with that in the Fermi-liquid-like phase, which shows no qualitative difference and indicates that the spin modes remain gapless at $\delta = 1/3$ [see Appendix~\ref{app:spin}].
These results of correlation functions describe this state as a possible precursor of a superconducting state, such as the pseudogap state.

We further examine the $1/3$ doping on the YC8 cylinder, but find that the character of hole pairing disappears. 
The magnitudes of the squared single-particle correlation and pairing correlation are again comparable [Figs.~\ref{033}(f)-(g)]. 
This indicates that the state near $1/3$ doping in the YC6 system may have a system size dependence, which requires further study.

\section{Summary and discussion}
Using DMRG calculations, we have investigated the kagome-lattice $t$-$J$ model at $\delta = 0.027-0.36$, which extends the doping range of previous DMRG studies~\cite{kagome-tJ-Jiang-2017,kagome-tJlike-PCheng-2021}. 
On the YC6 cylinder, the system shows a transition from CDW states to a Fermi-liquid-like phase at $\delta \approx 0.15$. 
In this metallic phase, the charge density oscillation is suppressed, and the measured correlation functions including single-particle, spin-spin, density-density, and pairing correlations all decay algebraically.
Furthermore, this state is characterized by the existence of gapless modes and the absence of hole pairing.
On the wider YC8 cylinder, the bond-dimension extrapolated ($D \rightarrow \infty$) correlation functions also support such a Fermi-liquid-like state, suggesting its stability with increasing system size.

Remarkably, this state with algebraic single-particle correlation is distinct from the non-Fermi liquid state found in the PESS simulation~\cite{kagome-tJ-GuZC-2024}, which shows an exponential decay of single-particle and spin correlations~\cite{law2017}.
While the phase transition from a Wigner crystal to a Fermi liquid with increasing electron density is conventional in the electron gas~\cite{Wigner-1934,2D-electron-gas,3D-electron-gas}, a transition from a holon Wigner crystal with fractionalization to a Fermi liquid may be highly non-trivial, which has been intensively studied in related models of the bilayer transition metal dichalcogenide materials~\cite{musser2022,mussuer2022_2}.
It would be very interesting to clarify the nature of this Fermi-liquid-like phase, e.g., a conventional Fermi liquid or a fractionalized Fermi liquid, and the transition properties in the melting of the holon Wigner crystal.
We leave these open questions to future study.

Near $1/3$ doping on the YC6 cylinder, the state with an exponential decay of single-particle correlation may be considered as a precursor of a superconducting state, similar to the pseudogap state.
This state may have a potential connection with the pair density wave state found in the PESS simulation at $\delta \in [0.32, 1/3]$~\cite{kagome-tJ-GuZC-2024}.
Nevertheless, due to the size dependence observed in the YC6 and YC8 systems, this doping range needs further studies on larger system size.
In addition, considering additional interactions such as  spin chiral interaction, flux, or density-density attraction may be helpful for stabilizing a superconducting phase near $1/3$ doping.

\begin{acknowledgments}
X.~Y.~J. and S.~S.~G. were supported by the National Natural Science Foundation of China (No. 12274014), the Special Project in Key Areas for Universities in Guangdong Province (No. 2023ZDZX3054), and the Dongguan Key Laboratory of Artificial Intelligence Design for Advanced Materials. D.~N.~S. was supported by the U.S. Department of Energy, Office of Basic Energy Sciences under Grant No. DE-FG02-06ER46305 for DMRG studies of unconventional superconductivity.
The computational resources were supported by the SongShan Lake HPC Center (SSL-HPC) at Great Bay University (X. Y. J. and S. S. G.). The numerical simulation was in part supported by the US National Science Foundation instrument grant DMR-2406524 (D.~N.~S.).

\end{acknowledgments}

\appendix
\section{Averaged charge density per unit cell}
\label{app:cdw}

\begin{figure}[h] 
	\centering
	\begin{subfigure}[b]{0.48\textwidth}
		\includegraphics[width=\textwidth]{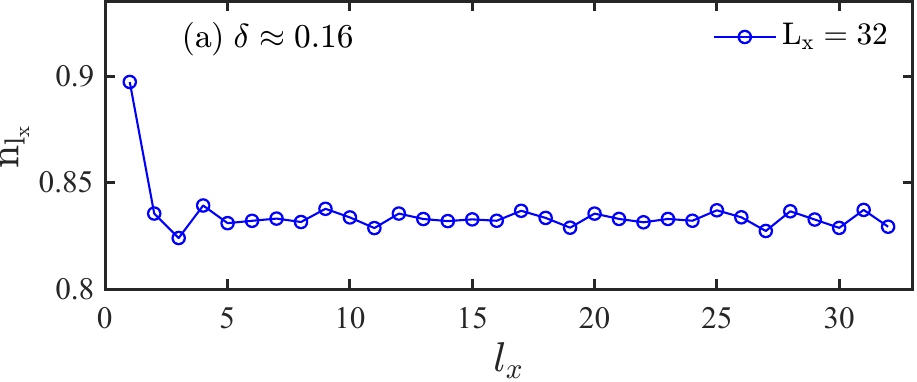} 
	\end{subfigure}
	\begin{subfigure}[b]{0.48\textwidth}
		\includegraphics[width=\textwidth]{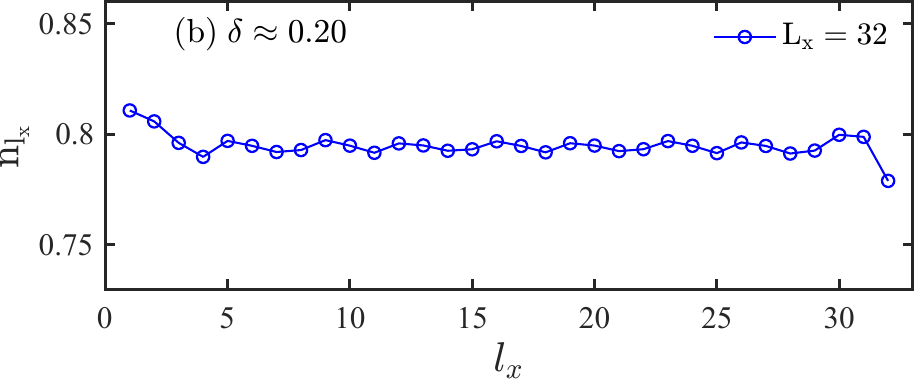} 
    \end{subfigure}
	\begin{subfigure}[b]{0.48\textwidth}
		\includegraphics[width=\textwidth]{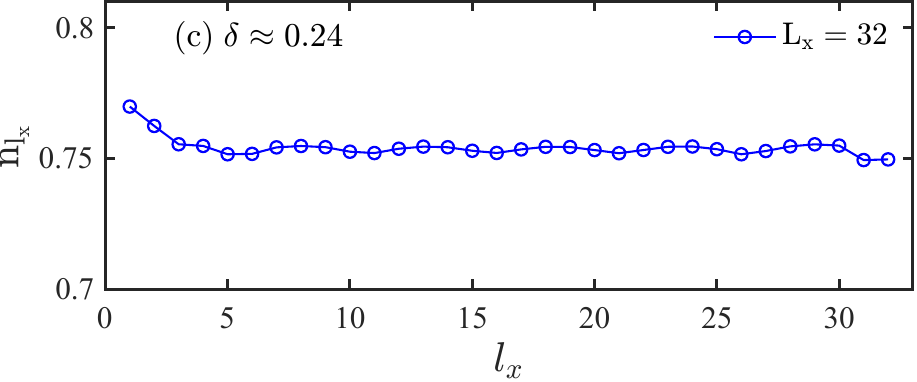} 
    \end{subfigure}
	\caption{\justifying Unit-cell-averaged charge density profiles on the YC6 cylinder with $L_x=32$. The averaged charge density of the unit cell in each column $n_{l_x}$ is defined as $n_{l_x} = \frac{1}{3L_y} \sum_{l_y=1}^{L_y} \sum_{i=1}^{3} \langle \hat{n}_{l_x,l_y,i} \rangle $, where $l_x$ ($l_y$) is the unit cell index along the ${\bf e}_1$ (${\bf e}_2$) direction and $i$ denotes the three sites in each unit cell. (a)-(c) $n_{l_x}$ in the Fermi-liquid-like phase at $\delta \approx 0.16$, $0.20$, and $0.24$, respectively.}
    \label{charge_unit}
\end{figure}

In the main text, we have presented the site-resolved charge density distribution. 
In this section, we further show the averaged charge density for each unit cell in Fig.~\ref{charge_unit}, which illustrates the variation of the charge density across different unit cells.
The average charge density of the unit cell in each column $n_{l_x}$ is defined as $n_{l_x} = \frac{1}{3L_y} \sum_{l_y=1}^{L_y} \sum_{i=1}^{3} \langle \hat{n}_{l_x,l_y,i} \rangle $, where $l_x$ ($l_y$) is the unit cell index along the ${\bf e}_1$ (${\bf e}_2$) direction and $i$ denotes the three sites in each unit cell.
It is evident that for the doping range $\delta = 1/6-1/4$, the unit-cell-averaged charge density is nearly uniform along the ${\bf e}_1$ direction.

\section{Extrapolation of entanglement entropy with growing bond dimension}
\label{app:entropy}

\begin{figure}[t]
	\centering
    \includegraphics[width=0.45\textwidth]{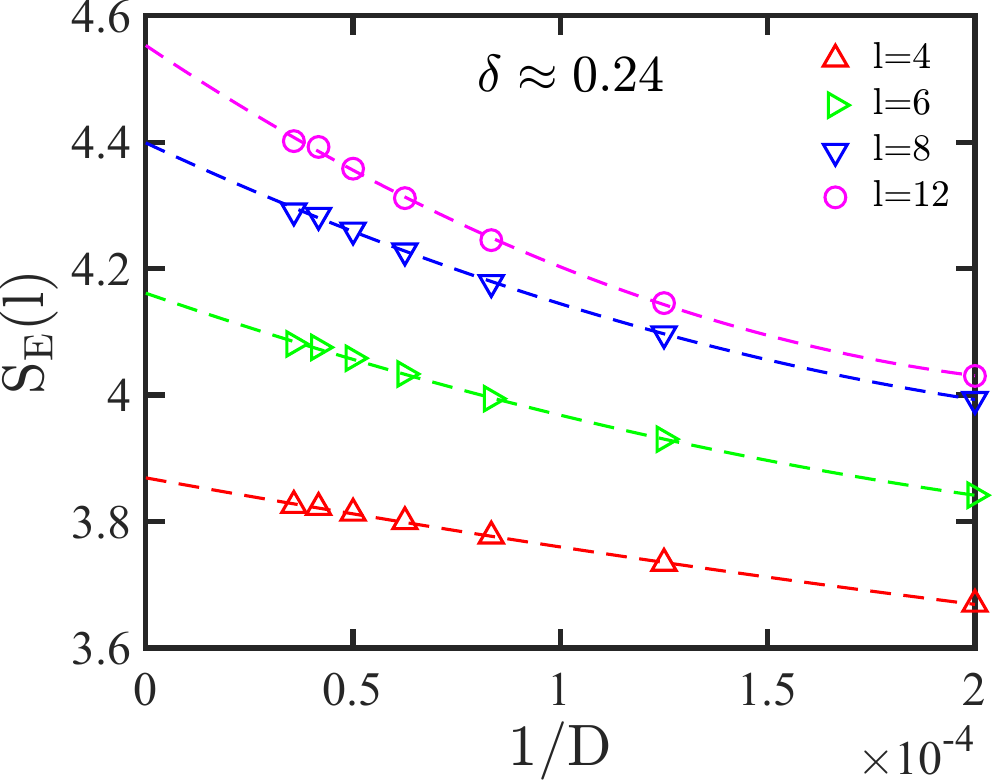} 
	\caption{\justifying Extrapolations of entanglement entropy $S_{E}(l)$ versus bond dimension at $\delta \approx 0.24$ on the YC6 cylinder with $L_x = 32$. The $SU(2)$ bond dimension $D$ ranges from $8000$ to $28000$. The label $l$ denotes the unit cell index along the ${\bf e}_1$ direction. For each given $l$, the entropy obtained by different bond dimensions are extrapolated by the polynomial function $C(1/D) = C(0) + a/D + b/D^2$.} 
    \label{Extrapolate}
\end{figure}

DMRG simulations inevitably have the finite-bond-dimension effect. 
To reduce this effect, the extrapolated entropy is shown in the main text, and here we show the extrapolation detail.
We first obtain the entanglement entropy at different bond dimensions, and then perform a polynomial extrapolation for the data at different $D$ to extract the result in the infinite-$D$ limit.
We fit the data for a range of bond dimensions up to the largest $SU(2)$ bond dimensions $D = 28000$ (equivalent to about $84000$ $U(1)$ states). 
A typical example of data extrapolation is shown for $\delta \approx 0.24$ in Fig.~\ref{Extrapolate}. 
For each given $l$, the entanglement entropys obtained by seven different bond dimensions are extrapolated by the polynomial function $C(1/D) = C(0) + a/D + b/D^2$, where $C(0)$, $a$, and $b$ are determined by fitting the DMRG data. 
The obtained $C(0)$ is the entropy in the infinite-$D$ limit.

\section{Spin correlation function for $1/3$ doping on the YC6 cylinder}
\label{app:spin}

\begin{figure}[t]
	\centering
    \includegraphics[width=0.45\textwidth]{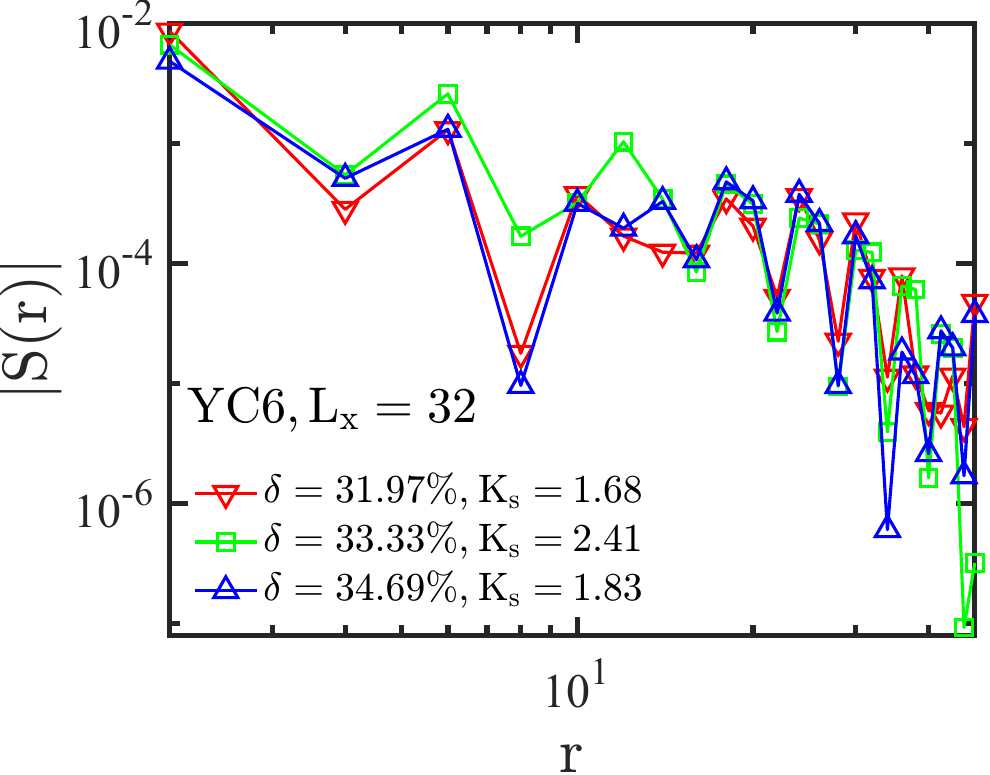} 
	\caption{\justifying Spin correlation functions $S(r)$ for several doping levels around $\delta = 1/3$ on the YC6 cylinder with $L_x =32$.} 
    \label{app-spin}
\end{figure}

In this section, we compare the spin correlation functions near the doping level of $1/3$ on the YC6 cylinder. As illustrated in Fig.~\ref{app-spin}, the decay behavior of the spin correlation functions at various doping levels around $\delta = 1/3$ is consistent, displaying a power-law decay. 
While the single-particle correlation function becomes an exponential decay at $1/3$ doping, the spin correlation function remains the feature in the Fermi-liquid-like phase.

\bibliographystyle{apsrev4-2}
\bibliography{kagome}

\end{document}